\documentclass{ws-m3as}
\usepackage[T1]{fontenc}
\usepackage[super]{cite}
\usepackage{graphicx}
\usepackage{bm}
\usepackage{tikz}
\usepackage[percent]{overpic}
\usepackage{makecell}
\usetikzlibrary{arrows.meta,positioning,calc}
\usepackage{booktabs,array}
\usepackage{pgfplots}
\usepackage{tabularx}
\usepackage{multirow}
\usepackage{subcaption}
\usepackage{float}
\usepackage{xcolor}
\pgfplotsset{compat=1.17}
\usetikzlibrary{arrows.meta,positioning,calc,shapes.geometric}

\usepackage[verbose,hypertexnames=false]{hyperref}
\hypersetup{
  colorlinks=false,
  allbordercolors=blue,
  pdfborderstyle={/S/U/W 1}
}

\newcommand{\SmA}{\mathrm{SmA}}
\newcommand{\SmC}{\mathrm{SmC}}
\newcommand{\NF}{\mathrm{N_F}}
\newcommand{\SmAF}{\mathrm{SmA_F}}
\newcommand{\SmCF}{\mathrm{SmC_F}}

\newcommand{\by}{\mathbf{y}}
\newcommand{\bg}{\mathbf{g}}

\renewcommand{\subsectionfont}{%
  \bfseries\boldmath\upshape
  \rightskip2pc plus1fill\relax
}

\renewcommand{\subsubsectionfont}{%
  \bfseries\boldmath\upshape
  \rightskip2pc plus1fill\relax
}
\begin{document}
\setlength{\paperwidth}{210mm}
\setlength{\paperheight}{297mm}
\markboth{S. Guan, Y. Han and L. Zhang}{Modular Landau Analysis}
\catchline{}{}{}{}{}

\title{Modular Landau Analysis of Coupled Phase Transitions: Application to Ferroelectric Liquid Crystals}

\author{Sisi Guan}
\address{Beijing International Center for Mathematical Research, Peking University\\
No. 5 Yiheyuan Road, Haidian District, Beijing 100871, China\\
sisiguan98@bicmr.pku.edu.cn}

\author{Yucen Han}
\address{Center for Applied Mathematics, Renmin University of China\\
No. 59 Zhongguancun Street, Haidian District, Beijing 100872, China\\
yucen.han@ruc.edu.cn}

\author{Lei Zhang}
\address{School of Mathematical Sciences, Beijing International Center for Mathematical Research,\\
Center for Quantitative Biology, Center for Machine Learning Research,\\
Institute for Artificial Intelligence, Peking University\\
No. 5 Yiheyuan Road, Haidian District, Beijing 100871, China\\
zhangl@math.pku.edu.cn}
\maketitle

\begin{abstract}
In systems with coupled order parameters, the relaxation of existing order can modify subsequent transitions, while cooperation and competition between emerging modes can produce complex phase sequences. We develop a modular Landau analysis that locally relaxes noncritical parent-phase variables to obtain effective one- and two-mode free energies. The resulting coefficients quantify relaxation-induced changes in transition characteristics and mode coupling, including changes in transition order for even-symmetry modes. The resulting modules are assembled successively along parent branches, and the equilibrium phase sequence is selected by comparing the fully relaxed free energies of the admissible branches. We apply the framework to ferroelectric liquid crystals, where orientational, polar, smectic, and tilt orders interact and give rise to multiple competing pathways, using a uniaxial model with longitudinal polarization. We fit the model to the measurements reported by Strachan \textit{et al.} [\textit{Adv. Sci.} \textbf{12}(3), 2409754 (2025)]. A single thermodynamic parameter set reproduces the observed transition temperatures, transition types, and phase sequence, while capturing the main birefringence trend. Smectic layer spacing is reconstructed separately through a fitted observation map based on the calculated equilibrium tilt.
\end{abstract}


\ccode{2020 Mathematics Subject Classification: Primary 82B26; Secondary 82D30}

\section{Introduction}\label{sec:introduction}

Phase transitions are qualitative reorganizations of matter that occur when changes in temperature, pressure, composition, or external fields alter the symmetry or character of an equilibrium state \cite{Landau1937,ChaikinLubensky1995}. Familiar examples include melting and vaporization, magnetic and ferroelectric ordering, superconductivity, and structural transformations \cite{Ising1925,Devonshire1949,GinzburgLandau1950,Cowley1980}. According to their thermodynamic behavior, phase transitions are commonly classified as first-order or continuous, with tricritical points separating these regimes; they may also be distinguished by the symmetry and order parameter involved \cite{Landau1937,ChaikinLubensky1995}. Theoretical descriptions range from microscopic statistical models to phenomenological free-energy theories and renormalization-group approaches \cite{Ising1925,WilsonFisher1972}. Among these approaches, Landau theory represents the state of a system using a small set of order parameters and expands an effective free energy in symmetry-allowed terms. It therefore provides analytically tractable criteria for transition type, phase stability, and coupling between different ordering modes without requiring a detailed microscopic description \cite{Landau1937,ChaikinLubensky1995,Cowley1980}.

For a single ordering mode, the local Landau expansion is constrained by symmetry and thermodynamic stability, giving continuous, tricritical, or first-order behavior depending on the allowed terms and effective coefficients \cite{Landau1937,ChaikinLubensky1995,Cowley1980,LawrieSarbach1984,ToledanoToledano1987}. When several order parameters are coupled, however, an emerging mode is generally coupled to variables already present in the parent phase, whose relaxation renormalizes the effective Landau coefficients \cite{Cowley1980,ToledanoToledano1987,StokesHatch1991,Toledano2012}. This renormalization can shift coexistence and metastability limits, modify the order-parameter jump, and, when the effective quartic coefficient changes sign, alter the transition type. Although such relaxation effects are well established, they are not always retained explicitly in reduced descriptions, where the response of the parent variables to the emerging mode may be omitted \cite{ChenLubensky1976,Mukherjee2010}.

Additional complexity arises when more than one ordering mode is relevant near the same parent phase. Their coupling may favor cooperative mixed ordering or competition between distinct ordered states, producing sequential transitions, mixed phases, and alternative transition pathways \cite{ChaikinLubensky1995,Cowley1980,ToledanoToledano1987}. When several locally stable branches coexist, local stability alone does not determine the equilibrium phase; the competing branches must instead be compared through their fully relaxed free energies. Related studies have investigated competing equilibria, free-energy landscapes, and transition pathways in specific models \cite{Biscari2007,PajakOsipov2013,HanHuZhangZhang2019,ShiHanZhang2022,ShiHan2024Multistability,ShiHanYin2023CriticalPoints}. The present work complements these studies by developing a reusable reduction framework that organizes local transition mechanisms and global branch selection.

Landau theory has been widely applied to phase-transition problems involving coupled order parameters. In structural and ferroelectric systems, Landau--Devonshire theories and related variational formulations describe symmetry breaking, polarization, and transitions between ordered phases \cite{Devonshire1949,Cowley1980,VanderbiltCohen2001,LiCrossChen2005,DuLiZhang2020}. In multiferroic systems, phenomenological free energies describe the coupling between magnetic and electric polarization orders \cite{Mostovoy2006}. Coupled Landau expansions are also used to study competing orders in superconductors, including the coexistence and competition between magnetism and superconductivity \cite{FernandesSchmalian2010}, as well as bicritical and tetracritical structures generated by coupled order parameters \cite{BruceAharony1975}. Landau free-energy models have also been used to study transitions between distinct ordered phases \cite{Yin2021TransitionPathways,ZhangEtAl2026LandauBrazovskii}. These examples illustrate the broad applicability of Landau theory to systems with multiple interacting order parameters.

Motivated by these considerations, we develop a modular Landau framework for coupled phase transitions. Starting from a locally stable parent phase, the noncritical parent variables are relaxed with respect to the emerging ordering modes, producing effective one- and two-mode reductions with systematically renormalized Landau coefficients. For the one-mode \(2\!-\!3\!-\!4\) and \(2\!-\!4\!-\!6\) models, comparison with the corresponding frozen-parent descriptions makes explicit how relaxation modifies first-order transition characteristics and, in the symmetry-preserving case, can even change the transition character. The same relaxation principle extends to coupled ordering modes through a renormalized quartic form. The selected daughter branch then becomes the parent for the next transition, so that complete equilibrium phase sequences are assembled from local transition modules and selected by comparison of fully relaxed branch free energies.

Liquid crystals provide a particularly suitable setting for this construction because orientational, polar, smectic-layering, and tilt orders can appear successively and may either cooperate or compete \cite{deGennesProst1993,Chandrasekhar1992,Larson1999,Hamley2010,WangZhangZhang2021}. Landau-type models have long been used to describe the conventional \(I\), \(N\), \(\SmA\), and \(\SmC\) phases \cite{ChenLubensky1976,Biscari2007,deGennes1971,deGennes1972,Izzo2020}, with further studies of smectic layering, tilt, and reduced nematic equilibria \cite{ShiHan2025,XiaHan2024,HanShiZhang2023Prisms}. The discovery of the ferroelectric \(N_F\), \(\SmA_F\), and \(\SmC_F\) phases further enlarges this landscape by introducing coupled polar, orientational, smectic, and tilt orders, together with a variety of sequential and direct ordering pathways \cite{Mandle2017,Nishikawa2017,Chen2020,Kikuchi2022,Song2022,Chen2022,Kikuchi2024,Strachan2025,Weglowska2025}. This makes ferroelectric liquid crystals a natural test case for the modular framework. We therefore construct a symmetry-consistent \(Q+P+\psi\) free energy and use a unified thermodynamic framework to compare competing cooling pathways. For this application, we adopt a homogeneous uniaxial ansatz with longitudinal polarization \(P=pn\) and fixed smectic wave number \(q=q_0\). Fitting to the measurements of Strachan \textit{et al.}~\cite{Strachan2025} reproduces the observed phase sequence, transition types, and the main orientational and polar trends.

The remainder of this paper is organized as follows. Section~\ref{sec:universal-landau-reductions} develops the general modular Landau framework, including parent-branch relaxation, effective one- and two-mode reductions, the influence of relaxation on their effective coefficients and transition behavior, and modular phase selection. Section~\ref{sec:liquid-crystal-application} applies the framework to ferroelectric liquid crystals and fits the model to a representative experiment, showing good agreement with the observed phase behavior. Finally, Section~\ref{sec:conclusion} summarizes the main results and discusses possible extensions.
\section{Modular Phase Transition Analysis in Landau Theory}
\label{sec:universal-landau-reductions}

\begin{figure}[H]
    \centering
    \includegraphics[width=0.82\linewidth]{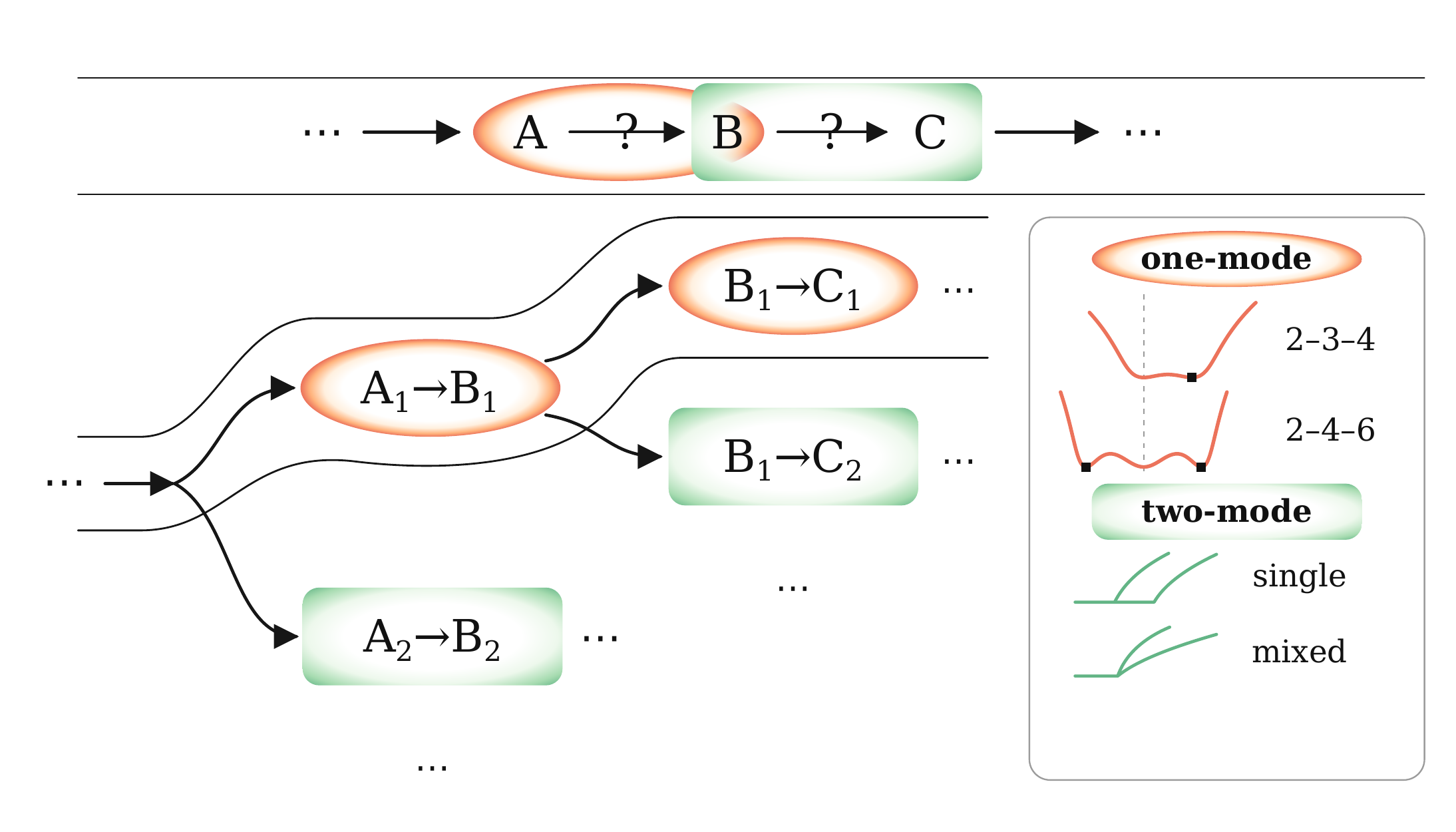}
    \caption{Modular construction of a phase sequence.}
    \label{fig:modular}
\end{figure}

Within the modular framework, a phase sequence is decomposed into successive local transition modules, each defined relative to a locally stable parent phase and classified by the number and symmetry of the emerging ordering modes. For example, consider a segment $\cdots \to A \to B \to C \to \cdots$. The transitions \(A_i\to B_j\) and \(B_j\to C_k\) are analyzed as separate local problems. Once a daughter branch is selected, it becomes the parent branch for the subsequent transition. This construction is illustrated schematically in Fig.~\ref{fig:modular}.

In Fig.~\ref{fig:modular}, one-mode reductions are represented by red ellipses and two-mode reductions by green squares. The one-mode reductions distinguish asymmetric from symmetry-preserving ordering and lead, respectively, to the \(2\!-\!3\!-\!4\) and \(2\!-\!4\!-\!6\) models, retaining quadratic, cubic, and quartic terms in the former and quadratic, quartic, and sextic terms in the latter. When two ordering modes are simultaneously relevant, they are retained within a common two-mode reduction so that single-mode ordering (producing sequential transitions), mixed ordering (two order parameters arising simultaneously), and competition between the associated branches can be treated within the same local problem. The local reductions generate candidate daughter branches. For example, after selecting \(B_1\) as the parent branch, there may be multiple candidate daughters, \(C_1,C_2,C_3,\ldots\), associated with different one-mode or two-mode ordering channels. Higher-mode reductions may also arise in principle, but are not considered here. The equilibrium branch is selected by comparing the fully relaxed free energies of the candidate branches. Among the resulting candidate phase sequences, a sequence such as \(\cdots\to A_1\to B_1\to C_1\to\cdots\) is selected according to the full free energy.


\subsection{Parent branches and relaxation}
\label{subsec:parent-branch-relaxation}

Let $\alpha$ denote a candidate parent phase, and let $\by\in\mathbb R^m$ denote the noncritical variables parametrizing that phase.  Let $\bm\xi$ denote the ordering coordinates that vanish in the parent phase. We write the local free energy as $f(\by,\bm\xi;T)$ and define its restriction to the parent phase by
\[
    f_\alpha(\by;T):=f(\by,\bm 0;T).
\]
The locally stable parent branch $\bar{\by}_\alpha(T)$ satisfies
\begin{equation}
    \nabla_{\by}f_\alpha(\bar{\by}_\alpha;T)=0,\qquad H_\alpha(T):=D_{\by}^2 f_\alpha(\bar{\by}_\alpha;T)\succ0.\label{eq:parent-branch}
\end{equation}
Thus the parent variables remain locally nondegenerate within the parent phase. In contrast, the ordering coordinates $\bm\xi$ describe departures from that phase and may become critical at a transition. We assume that $f$ admits a local Landau expansion about $(\bar{\by}_\alpha,\bm 0)$, with $\nabla_{\bm\xi}f(\bar{\by}_\alpha,\bm0;T)=0$, so that the leading stability of the parent state in the ordering directions is determined by the corresponding quadratic terms.

Since $H_\alpha(T)$ is nonsingular, the implicit-function theorem gives, for sufficiently small $\bm\xi$, a locally unique relaxed branch $\by=\by_\alpha(\bm\xi;T)$ with $\by_\alpha(\bm0;T)=\bar{\by}_\alpha(T)$ satisfying
\begin{equation}
    \nabla_{\by}f\bigl(\by_\alpha(\bm\xi;T),\bm\xi;T\bigr)=0,\qquad f_{\rm eff}(\bm\xi;T):=f\bigl(\by_\alpha(\bm\xi;T),\bm\xi;T\bigr).\label{eq:general-parent-relaxation}
\end{equation}
Unless an argument is displayed explicitly, every coefficient and $\by$-derivative below is evaluated at $\by=\bar{\by}_\alpha(T)$.  The same construction is applied whenever the selected daughter phase becomes the parent for a subsequent transition.

\subsection{One-mode reductions}
\label{subsec:one-mode}
For a single ordering mode, the lowest polynomial order in a local analytic Landau expansion is constrained by symmetry, stability, and the local topology of stationary branches. A first-order transition requires coexistence of two minima separated by an intermediate maximum, and hence the derivative of the free energy must possess at least three real zeros. Therefore, a quartic expansion is the minimal stable description. When odd powers are allowed, this gives the asymmetric \(\mathit{2\!-\!3\!-\!4}\) \textit{model},
\[
    f(\xi)=\frac12 r\xi^2-\frac13 b\xi^3+\frac14u\xi^4,\qquad u>0 .
\]
If the ordering coordinate has an even symmetry \(f(\xi)=f(-\xi)\), the quartic form $f(\xi)=\frac12 r\xi^2+\frac14u\xi^4$ only describes a continuous symmetry-breaking transition. A symmetric first-order transition requires an additional pair of stationary branches, and therefore the minimal even expansion is the \(\mathit{2\!-\!4\!-\!6}\) \textit{model},
\[
    f(\xi)=\frac12r\xi^2+\frac14u\xi^4+\frac16v\xi^6,\qquad v>0 .
\]
This model contains continuous, tricritical, and first-order transitions depending on the signs and degeneracies of the quadratic and quartic coefficients \cite{Landau1937,ChaikinLubensky1995,LawrieSarbach1984,ToledanoToledano1987,NishimoriOrtiz2011}. Higher-order terms are required only when additional stationary branches, more complicated phase topologies, or lower-symmetry states need to be represented \cite{VanderbiltCohen2001,LiCrossChen2005}.

The quadratic coefficient \(A_\xi\) may generally depend on external control parameters. Since we consider temperature-driven transitions, we retain only its leading temperature dependence near the instability, while treating higher-order coefficients and coupling parameters as approximately constant over the relevant temperature range. The same convention is used in Section~\ref{sec:liquid-crystal-application}.

\subsubsection{Asymmetric \texorpdfstring{\(2\!-\!3\!-\!4\)}{2-3-4} model}
\label{subsec:234}

If symmetry permits odd powers of the ordering coordinate \(\xi\), the local free energy near the parent branch can be expanded as
\begin{equation}
    f(\by,\xi;T)=\,f_\alpha(\by;T)+\frac12 A_\xi(\by,T)\xi^2-\frac13 \Gamma_\xi(\by,T)\xi^3+\frac14 B_\xi(\by,T)\xi^4+O(\xi^5).\label{eq:general-234-original-energy}
\end{equation}
Here \(\xi=0\) corresponds to the parent phase. As in Section~\ref{subsec:parent-branch-relaxation}, all coefficients and \(\by\)-derivatives below are evaluated at \(\by=\overline{\by}_\alpha(T)\). Define $\bg:=\nabla_{\by}A_\xi(\overline{\by}_\alpha,T).$

Allowing the parent variables to relax with the ordering coordinate, write
\begin{equation}
    \by(\xi;T)=\overline{\by}_\alpha(T)+\by_1\xi+\by_2\xi^2+O(\xi^3).\label{eq:234-y-expansion}
\end{equation}
Substituting \eqref{eq:234-y-expansion} into $\nabla_{\by}f(\by(\xi;T),\xi;T)=0$ and expanding about \(\overline{\by}_\alpha(T)\) gives
\begin{align*}
    0=H_\alpha\by_1\,\xi+\left[H_\alpha\by_2+\frac12D_{\by}^3f_\alpha[\by_1,\by_1]+\frac12\bg\right]\xi^2+O(\xi^3).
\end{align*}
Matching powers of \(\xi\) first gives $H_\alpha\by_1=0$. Since \(H_\alpha\succ0\), \(\by_1=0\). The \(O(\xi^2)\) equation then reduces to $H_\alpha\by_2+\frac12\bg=0$, so that $\by_2=-\frac12H_\alpha^{-1}\bg$. Hence
\begin{equation}\label{eq:234-y-expansion-1}
    \by(\xi;T)=\overline{\by}_\alpha(T)-\frac12H_\alpha^{-1}\bg\,\xi^2+O(\xi^3).
\end{equation}

The corresponding relaxed free-energy difference from the parent branch is
\begin{equation*}
    \Delta f_{\mathrm{eff}}(\xi;T):=f_{\mathrm{eff}}(\xi;T)-f_{\mathrm{eff}}(0;T)=f\bigl(\by(\xi;T),\xi;T\bigr)-f_\alpha\bigl(\overline{\by}_\alpha(T);T\bigr).
\end{equation*}
Using \eqref{eq:234-y-expansion-1} and expanding \eqref{eq:general-234-original-energy} about \(\overline{\by}_\alpha(T)\), we obtain
\begin{align*}
    \Delta f_{\mathrm{eff}}(\xi;T)
    =\frac12A_\xi\xi^2-\frac13\Gamma_\xi\xi^3+\frac14\left(B_\xi-\frac12\bg^TH_\alpha^{-1}\bg\right)\xi^4+O(\xi^5).
\end{align*}

Truncating at quartic order gives the reduced \(2\!-\!3\!-\!4\) model
\begin{equation}
    \Delta f_{234}(\xi;T)=\frac12r_\xi\xi^2-\frac13b_\xi\xi^3+\frac14u_\xi\xi^4,\qquad u_\xi>0,\label{eq:234-reduced-energy}
\end{equation}
with
\begin{equation}
    r_\xi=A_\xi,\qquad b_\xi=\Gamma_\xi,\qquad u_\xi=B_\xi-\frac12\bg^TH_\alpha^{-1}\bg.\label{eq:general-234-effective-coefficients}
\end{equation}
Since \(H_\alpha\succ0\), $\bg^TH_\alpha^{-1}\bg\geq0$, and hence \(u_\xi\leq B_\xi\). Thus relaxation of the parent variables lowers the effective quartic coefficient relative to the frozen-parent value.

If \(b_\xi\neq0\), the orientation of \(\xi\) may be chosen so that \(b_\xi>0\); the case \(b_\xi<0\) follows from \(\xi\mapsto-\xi\). The stationary points of \eqref{eq:234-reduced-energy} satisfy
\begin{equation*}
    \frac{\partial\Delta f_{234}}{\partial\xi}=\xi\left(r_\xi-b_\xi\xi+u_\xi\xi^2\right)=0.
\end{equation*}
Besides the parent solution \(\xi=0\), the nonzero stationary branches are
\begin{equation}
    \xi_\pm=\frac{b_\xi\pm\sqrt{b_\xi^2-4u_\xi r_\xi}}{2u_\xi}.\label{eq:234-nonzero-branches}
\end{equation}
For $0<r_\xi<b_\xi^2/(4u_\xi)$, the branch \(\xi_+\) is a local minimum and \(\xi_-\) is the intervening local maximum.


\paragraph{Ordered-branch spinodal.} 
In Landau theory, a spinodal is the limit of metastability at which a local minimum coalesces with an adjacent unstable stationary point \cite{Landau1937,ChaikinLubensky1995}. The nonzero stationary branches first appear when the discriminant in \eqref{eq:234-nonzero-branches} vanishes: $b_\xi^2-4u_\xi r_\xi=0$. Thus the ordered-branch spinodal \(T_{\rm sp}\) satisfies
\begin{equation*}
    r_\xi(T_{\rm sp})=\frac{b_\xi^2(T_{\rm sp})}{4u_\xi(T_{\rm sp})},\qquad\xi_{\rm sp}=\frac{b_\xi(T_{\rm sp})}{2u_\xi(T_{\rm sp})}.
\end{equation*}

\paragraph{Thermodynamic coexistence.}
At coexistence, the nonzero ordered minimum is stationary and has the same free energy as the parent state:
\begin{equation*}
    \partial_\xi\Delta f_{234}(\xi_{\rm coex};T_{\rm coex})=0,\qquad\Delta f_{234}(\xi_{\rm coex};T_{\rm coex})=0.
\end{equation*}
For \(\xi_{\rm coex}\neq0\), the first condition gives $r_\xi=b_\xi\xi_{\rm coex}-u_\xi\xi_{\rm coex}^2$. Substituting this relation into the second condition yields $0=b_\xi\xi_{\rm coex}^3/6-u_\xi\xi_{\rm coex}^4/4$, and therefore
\begin{equation*}
    \xi_{\rm coex}=\frac{2b_\xi(T_{\rm coex})}{3u_\xi(T_{\rm coex})},\qquad r_\xi(T_{\rm coex})=\frac{2b_\xi^2(T_{\rm coex})}{9u_\xi(T_{\rm coex})}.
\end{equation*}
The corresponding order-parameter jump is
\[
    \xi_{\rm jump}=|\xi_{\rm coex}|=\frac{2b_\xi(T_{\rm coex})}{3u_\xi(T_{\rm coex})}.
\]

\paragraph{Parent-branch spinodal.}
The parent branch loses local stability when its curvature at \(\xi=0\) vanishes: 
\[
    \left.\frac{\partial^2\Delta f_{234}}{\partial\xi^2}\right|_{\xi=0}=r_\xi.
\]
We denote this temperature by \(T_c\), so that $r_\xi(T_c)=0.$

For a temperature-driven cooling path with \(dr_\xi/dT>0\), assuming that \(b_\xi(T)\) and \(u_\xi(T)\) vary slowly near the transition, the characteristic temperatures satisfy $T_{\rm sp}>T_{\rm coex}>T_c.$

\paragraph{Effect of parent-branch relaxation.}
To quantify the effect of parent relaxation, compare the relaxed
coefficient \(u_\xi=B_\xi-\frac12\bg^T H_\alpha^{-1}\bg\) with the frozen-parent value \(u_\xi=B_\xi\). Under the same approximation that the higher-order coefficients vary slowly near the transition, the frozen-parent values are
\[
    r_\xi(T_{\rm sp}^{B})=\frac{b_\xi^2}{4B_\xi},\qquad r_\xi(T_{\rm coex}^{B})=\frac{2b_\xi^2}{9B_\xi},\qquad\xi_{\rm sp}^{B}=\frac{b_\xi}{2B_\xi},\qquad\xi_{\rm jump}^{B}=\xi_{\rm coex}^{B}=\frac{2b_\xi}{3B_\xi}.
\]
Since \(u_\xi\leq B_\xi\),
\[
    r_\xi(T_{\rm sp})\geq r_\xi(T_{\rm sp}^{B}), \qquad r_\xi(T_{\rm coex})\geq r_\xi(T_{\rm coex}^{B}),\qquad\xi_{\rm sp}\geq\xi_{\rm sp}^{B},\qquad\xi_{\rm coex}\geq\xi_{\rm coex}^{B}.
\]
For \(dr_\xi/dT>0\), this gives \(T_{\rm sp}\geq T_{\rm sp}^{B}\) and \(T_{\rm coex}\geq T_{\rm coex}^{B}\), while \(T_c\), determined by \(r_\xi(T_c)=0\), is unchanged. Thus parent-branch relaxation shifts the spinodal and coexistence temperatures upward and increases the order-parameter jump.

\begin{figure}[H]
    \centering
    \includegraphics[width=0.85\linewidth]{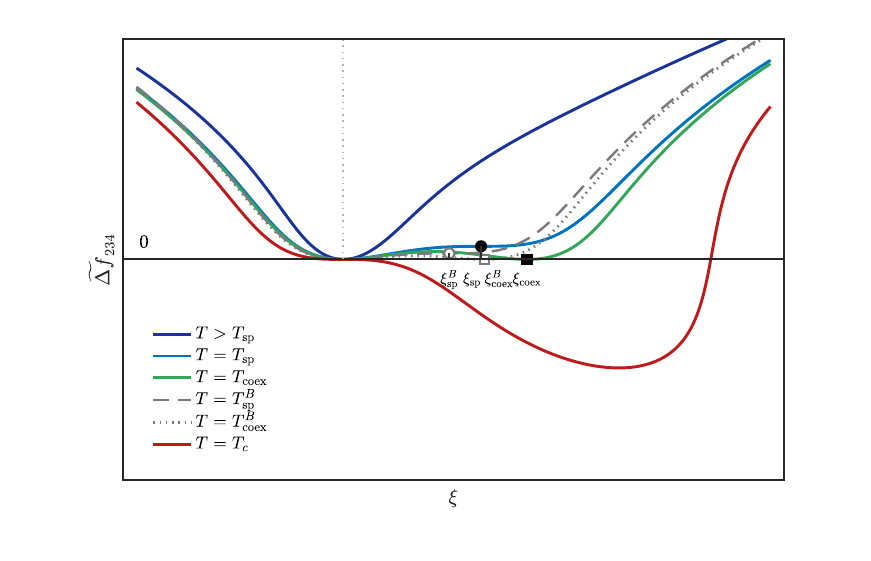}
    \vspace{-2.0em}
    \caption{Representative \(2\!-\!3\!-\!4\) free-energy landscapes during cooling. The gray dashed and dotted curves show the frozen-parent reference \(u_\xi=B_\xi\) at \(T=T_{\rm sp}^{B}\) and \(T=T_{\rm coex}^{B}\), respectively, with the corresponding \(\xi_{\rm sp}^{B}\) and \(\xi_{\rm coex}^{B}\) marked by the hollow circle and square.}
    \label{fig:234-first-order-landscape}
\end{figure}

Fig.~\ref{fig:234-first-order-landscape} summarizes this evolution and compares the relaxed landscapes with the corresponding frozen-parent spinodal and coexistence states.

\subsubsection{Symmetric \texorpdfstring{\(2\!-\!4\!-\!6\)}{2-4-6} model}
\label{subsec:246}

For an ordering coordinate invariant under \(\xi\mapsto-\xi\), the local free energy is even in \(\xi\). We write
\begin{equation}
    f(\by,\xi;T)=\,f_\alpha(\by;T)+\frac12A_\xi(\by,T)\xi^2+\frac14B_\xi(\by,T)\xi^4+\frac16C_\xi(\by,T)\xi^6+O(\xi^8).\label{eq:universal-original-energy-1}
\end{equation}
As before, all coefficients and \(\by\)-derivatives below are evaluated at \(\by=\overline{\by}_\alpha(T)\). Define
\begin{equation*}
    \bg:=\nabla_{\by}A_\xi,\qquad M:=D_{\by}^2A_\xi,\qquad \bm h:=\nabla_{\by}B_\xi,\qquad\mathcal T_\alpha:=D_{\by}^3f_\alpha,
\end{equation*}
where $\mathcal T_\alpha[\bm a,\bm b,\bm c]=\sum_{ijk}(\partial_{ijk}f_\alpha)a_i b_j c_k.$

Because the free energy is even in \(\xi\), the relaxed parent variables are also even in \(\xi\). We therefore write
\begin{equation}
    \by(\xi;T)=\overline{\by}_\alpha(T)+\by_2\xi^2+\by_4\xi^4+O(\xi^6).\label{eq:246-y-expansion}
\end{equation}
Substituting \eqref{eq:246-y-expansion} into $\nabla_{\by}f(\by(\xi;T),\xi;T)=0$ and collecting the \(O(\xi^2)\) terms gives $H_\alpha\by_2+\frac12\bg=0$. Since \(H_\alpha\succ0\),
\begin{equation*}
    \by_2=-\frac12H_\alpha^{-1}\bg,\qquad\by(\xi;T)=\overline{\by}_\alpha(T)-\frac12H_\alpha^{-1}\bg\,\xi^2+O(\xi^4).
\end{equation*}

As in the \(2\!-\!3\!-\!4\) reduction, define the relaxed energy relative to the parent branch by
\begin{equation*}
    \Delta f_{\mathrm{eff}}(\xi;T):=f_{\mathrm{eff}}(\xi;T)-f_{\mathrm{eff}}(0;T)=f\bigl(\by(\xi;T),\xi;T\bigr)-f_\alpha\bigl(\overline{\by}_\alpha(T);T\bigr).
\end{equation*}
Substituting \eqref{eq:246-y-expansion} into \eqref{eq:universal-original-energy-1} and expanding to sixth order gives
\begin{equation*}
    \Delta f_{\mathrm{eff}}(\xi;T)=\frac12r_\xi\xi^2+\frac14u_\xi\xi^4+\frac16v_\xi\xi^6+O(\xi^8),
\end{equation*}
where
\begin{equation}
    r_\xi=A_\xi,\qquad u_\xi=B_\xi-\frac12\bg^TH_\alpha^{-1}\bg,\label{eq:246-ru}
\end{equation}
and
\begin{equation}
    v_\xi=C_\xi+\mathcal T_\alpha[\by_2,\by_2,\by_2]+\frac32\by_2^TM\by_2+\frac32\bm h\cdot\by_2.\label{eq:246-v}
\end{equation}

Truncating at sixth order gives the reduced
\(2\!-\!4\!-\!6\) model
\begin{equation}
    \Delta f_{246}(\xi;T)=\frac12r_\xi\xi^2+\frac14u_\xi\xi^4+\frac16v_\xi\xi^6,\qquad v_\xi>0.\label{eq:reduced-energy-1}
\end{equation}
The effective quartic coefficient contains the same relaxation correction as in \eqref{eq:general-234-effective-coefficients},
\[
    u_\xi=B_\xi-\frac12\bg^T H_\alpha^{-1}\bg\leq B_\xi.
\]
Unlike the asymmetric case, the sign of \(u_\xi\) determines whether the symmetric transition is continuous, tricritical, or first order. Parent-branch relaxation can therefore move the system between these regimes, as discussed below.

The stationary equation of \eqref{eq:reduced-energy-1} is
\begin{equation}
    \frac{\partial\Delta f_{246}}{\partial\xi}=\xi\left(r_\xi+u_\xi\xi^2+v_\xi\xi^4\right)=0.\label{eq:246-stationary-equation}
\end{equation}
Introducing $x:=\xi^2\geq0$, the nonzero stationary states satisfy $v_\xi x^2+u_\xi x+r_\xi=0$, and hence
\begin{equation}
    x_\pm=\xi_\pm^2=\frac{-u_\xi\pm\sqrt{u_\xi^2-4v_\xi r_\xi}}{2v_\xi}.\label{eq:246-nonzero-branches}
\end{equation}
Let $x_{\rm eq}(T):=\xi_{\rm eq}^2(T)$ denote the globally minimizing equilibrium value of \(x=\xi^2\). The stationary structure of \eqref{eq:reduced-energy-1} gives three standard transition regimes, distinguished by the effective quartic coefficient $u_\xi$.

\paragraph{Continuous transition.}
Let $T_c$ be defined by $r_\xi(T_c)=0$. If $u_\xi(T_c)>0$, the ordered minimum emerges continuously from the parent branch. For \(r_\xi<0\), the nonzero branch in \eqref{eq:246-nonzero-branches} satisfies
\[
    \xi_{\rm eq}^2=\frac{-u_\xi+\sqrt{u_\xi^2-4v_\xi r_\xi}}{2v_\xi}=-\frac{r_\xi}{u_\xi}+O(r_\xi^2),
\]
where the last equality follows by expanding the square root for \(r_\xi\to0\) with \(u_\xi>0\). Hence $|\xi_{\rm eq}|\to 0$ as $T\uparrow T_c$, so that the transition is continuous. For a continuous transition, the ordered-branch spinodal, thermodynamic transition, and parent-branch spinodal coincide: $T_{\rm sp}=T_{\rm coex}=T_c$. 

\paragraph{Tricritical point.}
The boundary between the continuous and first-order regimes is the tricritical point, characterized by $r_\xi(T_{\rm tri})=u_\xi(T_{\rm tri})=0$ and $v_\xi(T_{\rm tri})>0$. For a nonzero stationary branch, \eqref{eq:246-stationary-equation} gives $r_\xi+u_\xi\xi^2+v_\xi\xi^4=0$. Let \(\tau=T-T_{\rm tri}\). Since \(r_\xi(T_{\rm tri})=u_\xi(T_{\rm tri})=0\) and \(v_\xi(T_{\rm tri})>0\), smoothness gives
\[
    r_\xi=O(\tau),\qquad u_\xi=O(\tau),\qquad v_\xi=O(1).
\]
Balancing the leading terms \(r_\xi\) and \(v_\xi\xi^4\) gives \(\xi^4=O(\tau)\), and hence \(u_\xi\xi^2=O(|\tau|^{3/2})\), which is of higher order. Therefore, $\xi_{\rm eq}^4\simeq-r_\xi/v_\xi$. For an ordinary temperature crossing with $r_\xi(T)\propto T-T_{\rm tri}$,
\[
    |\xi_{\rm eq}|\sim (T_{\rm tri}-T)^{1/4}.
\]

\paragraph{First-order transition.}
If $u_\xi<0$ and $v_\xi>0$, the reduced model can support a first-order transition. In this case the thermodynamic transition is determined by coexistence of the parent and ordered minima,
\[
    \Delta f_{246}(0;T_{\rm coex})=\Delta f_{246}(\xi_{\rm coex};T_{\rm coex}),\qquad \xi_{\rm coex}\neq0.
\]

The two nonzero stationary branches first appear when the discriminant in \eqref{eq:246-nonzero-branches} vanishes: $u_\xi^2-4v_\xi r_\xi=0$. Thus
\begin{equation*}
    r_\xi(T_{\mathrm{sp}})=\frac{u_\xi^2(T_{\mathrm{sp}})}{4v_\xi(T_{\mathrm{sp}})},\qquad\xi_{\mathrm{sp}}^2=-\frac{u_\xi(T_{\mathrm{sp}})}{2v_\xi(T_{\mathrm{sp}})}.
   \end{equation*}
At a nonzero stationary point, writing \(x=\xi^2\), \eqref{eq:246-stationary-equation} gives $r_\xi=-u_\xi x-v_\xi x^2$. Substitution into \eqref{eq:reduced-energy-1} yields
\[
    \Delta f_{246}=-\frac14u_\xi x^2-\frac13v_\xi x^3.
\]
At coexistence, \(\Delta f_{246}=0\), so $x_{\mathrm{coex}}=-3u_\xi/4v_\xi$. Hence
\begin{equation*}
    \xi_{\mathrm{coex}}^2=-\frac{3u_\xi(T_{\mathrm{coex}})}{4v_\xi(T_{\mathrm{coex}})},\qquad r_\xi(T_{\mathrm{coex}})=\frac{3u_\xi^2(T_{\mathrm{coex}})}{16v_\xi(T_{\mathrm{coex}})}.
\end{equation*}
The order-parameter jump at coexistence is therefore
\[
    |\xi_{\rm jump}|=\left(-\frac{3u_\xi(T_{\rm coex})}{4v_\xi(T_{\rm coex})}\right)^{1/2}.
\]
The parent branch remains locally stable at coexistence and becomes metastable upon further cooling. It loses local stability at \(T_c\),
defined by $r_\xi(T_c)=0$.

For a weakly first-order transition, \(u_\xi\) and \(v_\xi\) vary slowly over the metastability interval and may be approximated by their values at \(T_c\). Writing
\[
    r_\xi(T)=\dot r_\xi(T_c)(T-T_c)+o(T-T_c),\qquad\dot r_\xi(T_c)>0,
\]
and using the spinodal and coexistence conditions gives
\begin{equation*}
    T_{\mathrm{sp}}\simeq T_c+\frac{u_\xi^2(T_c)}{4\dot r_\xi(T_c)v_\xi(T_c)},\qquad T_{\mathrm{coex}}\simeq T_c+\frac{3u_\xi^2(T_c)}{16\dot r_\xi(T_c)v_\xi(T_c)}.
\end{equation*}
Hence $T_{\mathrm{sp}}>T_{\mathrm{coex}}>T_c.$ As \(|u_\xi(T_c)|\to0\), both the order-parameter jump and the metastability interval vanish, so the transition approaches the tricritical limit.

\par\medskip\noindent
\textbf{Effect of parent-branch relaxation.}\quad The preceding classification makes the effect of parent relaxation particularly clear. In the frozen-parent approximation, \(u_\xi^{B}=B_\xi\) and \(v_\xi^{B}=C_\xi\), whereas relaxation gives \(u_\xi\leq B_\xi\) together with the renormalized sextic coefficient \(v_\xi\) in \eqref{eq:246-v}. Hence, when \(B_\xi>0\), a sufficiently large relaxation correction can drive \(u_\xi\) through the tricritical value \(u_\xi=0\) and into the \(u_\xi<0\) regime. A transition that is continuous in the frozen-parent description can therefore become first order after relaxation, acquiring a finite order-parameter jump and a metastability interval \(T_c<T<T_{\rm sp}\).

If the frozen-parent model is already in the first-order regime (\(B_\xi<0\)), relaxation modifies the characteristic temperatures and jump through the combinations \(u_\xi^2/v_\xi\) and \(-u_\xi/v_\xi\). Their shifts, however, are not determined by \(u_\xi\leq B_\xi\) alone because the sextic coefficient is also renormalized.

\begin{figure}[t]
    \centering
    \begin{subfigure}[t]{0.485\linewidth}
        \centering
        \caption{Continuous transition, \(u_\xi>0\).}
        \includegraphics[width=\linewidth]{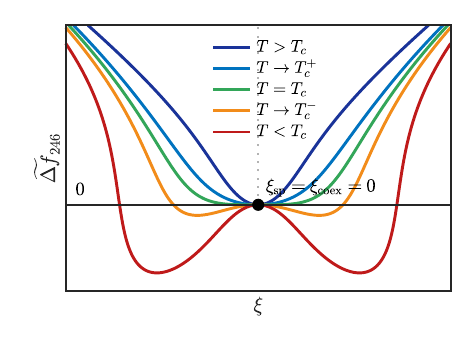}
        \label{fig:246-continuous}
    \end{subfigure}
    \hfill
    \begin{subfigure}[t]{0.485\linewidth}
        \centering
        \caption{First-order transition, \(u_\xi<0\), \(v_\xi>0\).}
        \includegraphics[width=\linewidth]{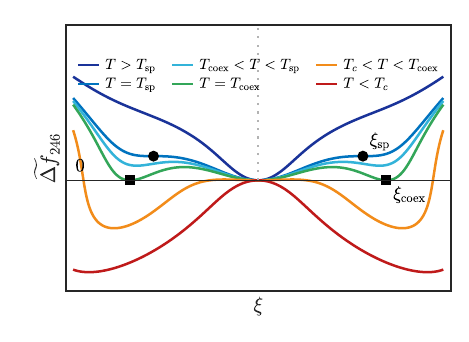}
        \label{fig:246-first-order}
    \end{subfigure}
    \caption{Representative \(2\!-\!4\!-\!6\) free-energy landscapes
    for (a) continuous and (b) first-order transitions.}
    \label{fig:246-landscapes}
\end{figure}

The continuous and first-order free-energy landscapes are compared in Fig.~\ref{fig:246-landscapes}. In the discussion below, \(T\to T_c^+\) denotes approach to \(T_c\) from the high-temperature side, whereas \(T\to T_c^-\) denotes approach from the low-temperature side; the superscripts indicate one-sided limits rather than distinct transition temperatures.

For \(u_\xi>0\), Fig.~\ref{fig:246-continuous} shows the continuous loss of stability of the parent minimum at \(\xi=0\). At \(T_c\), its quadratic curvature vanishes, and below \(T_c\) two symmetry-related ordered minima emerge continuously from the origin. Thus the ordered-branch spinodal, coexistence temperature, and parent-branch spinodal coincide at \(T_c\).

For \(u_\xi<0\) and \(v_\xi>0\), Fig.~\ref{fig:246-first-order} shows the first-order evolution of the free-energy landscape. A pair of nonzero ordered minima first appears at \(T_{\rm sp}\) while the parent minimum remains locally stable. At \(T_{\rm coex}\), the ordered and parent minima have equal free energy. Upon further cooling, the ordered minima become the global minima, while the parent minimum remains metastable until \(T_c\), where it loses local stability.

The first-order landscape is analogous to that of the asymmetric \(2\!-\!3\!-\!4\) model, but the symmetry is different. The cubic term in the \(2\!-\!3\!-\!4\) model selects a single ordered branch, whereas the even \(2\!-\!4\!-\!6\) model preserves the \(\xi\mapsto-\xi\) symmetry and produces the degenerate pair \(\pm\xi_{\rm eq}\).

\subsection{Two-mode competition and mixed ordering}\label{subsec:two-mode}
Let $\bm\xi=(\xi_1,\xi_2)$, $x_i:=|\xi_i|^2$, and $\bm x:=(x_1,x_2)^T$, with both ordering coordinates taken to be symmetry even. This even--even case is developed explicitly below, for which the reduced energy can be written entirely in terms of $x_1$ and $x_2$. If one mode is asymmetric, say $\xi_1$, while $\xi_2$ is even, one works instead with $(\xi_1,x_2)$ and retains the allowed odd and mixed terms, such as $\xi_1^3$, $\xi_1x_2$, and $\xi_1^2x_2$. If both modes are asymmetric, one works directly with $(\xi_1,\xi_2)$ and retains the symmetry-allowed cubic and quartic self- and mixed terms, such as $\xi_1^2\xi_2$ and $\xi_1\xi_2^2$. The parent variables are then relaxed through $\nabla_{\by}f=0$ exactly as below, and the resulting stationary branches are classified and compared by their fully relaxed free energies.

We focus on the even--even case because the coupled ordering modes considered in the applications below are of this form. Near a common parent branch, we write
\begin{equation}
    f(\by,\bm\xi;T)=\,f_\alpha(\by;T)+\frac12\bm A(\by,T)\cdot\bm x+\frac14\bm x^TU(\by,T)\bm x+f_6(\bm x;\by,T)+O(|\bm x|^4),\label{eq:universal-energy-2}
\end{equation}
where
\[
    \bm A=
    \begin{pmatrix}
        A_1\\ A_2
    \end{pmatrix},
    \qquad
    U=
    \begin{pmatrix}
        U_{11} & U_{12}\\
        U_{12} & U_{22}
    \end{pmatrix}.
\]
Here \(f_6\) is homogeneous of degree three in \(\bm x\), and is therefore sixth order in the ordering amplitudes.

As before, all coefficients and \(\by\)-derivatives below are evaluated at \(\by=\overline{\by}_\alpha(T)\), unless otherwise stated. Define the \(m\times2\) matrix
\begin{equation*}
    G:=\bigl[\nabla_{\by}A_1,\,\nabla_{\by}A_2\bigr].
\end{equation*}

Since \(\bm x=O(|\bm\xi|^2)\), write the relaxed parent variables as
\begin{equation}
    \by(\bm x;T)=\overline{\by}_\alpha(T)+Y_1\bm x+O(|\bm x|^2),\label{eq:universal-y-expansion-2}
\end{equation}
where \(Y_1\) is an \(m\times2\) matrix. Substituting \eqref{eq:universal-y-expansion-2} into $\nabla_{\by}f(\by(\bm x;T),\bm\xi;T)=0$ gives, to first order in \(\bm x\),
\[
    \left(H_\alpha Y_1+\frac12G\right)\bm x+O(|\bm x|^2)=0.
\]
Since this holds for arbitrary sufficiently small \(\bm x\), $H_\alpha Y_1+\frac12G=0$. Hence
\begin{equation*}
    Y_1=-\frac12H_\alpha^{-1}G,\qquad\by(\bm x;T)=\overline{\by}_\alpha(T)-\frac12H_\alpha^{-1}G\bm x+O(|\bm x|^2).
\end{equation*}

Define the relaxed energy relative to the parent branch by
\begin{equation*}
    \Delta f_{\mathrm{eff}}(\bm x;T):=f_{\mathrm{eff}}(\bm x;T)-f_{\mathrm{eff}}(\bm0;T)=f\bigl(\by(\bm x;T),\bm\xi;T\bigr)-f_\alpha\bigl(\overline{\by}_\alpha(T);T\bigr).
\end{equation*}
Substituting \eqref{eq:universal-y-expansion-2} into \eqref{eq:universal-energy-2} gives, through quadratic order in \(\bm x\),
\begin{equation*}
    \Delta f_{\mathrm{eff}}(\bm x;T)=\frac12\bm r\cdot\bm x+\frac14\bm x^TU_{\mathrm{eff}}\bm x+f_{6,\mathrm{eff}}(\bm x;T)+O(|\bm x|^4),
\end{equation*}
where
\begin{equation*}
    \bm r=\bm A(\overline{\by}_\alpha,T),\qquad U_{\mathrm{eff}}=U(\overline{\by}_\alpha,T)-\frac12G^TH_\alpha^{-1}G.
\end{equation*}
Since \(H_\alpha\succ0\), $G^TH_\alpha^{-1}G\succeq0.$ Thus relaxation of the common parent variables lowers the quartic form in every direction of the two-mode space, in the same way as in the one-mode reductions \eqref{eq:general-234-effective-coefficients} and \eqref{eq:246-ru}.

Write
\begin{equation*}
    \bm r=
    \begin{pmatrix}
        r_1\\ r_2
    \end{pmatrix},
    \qquad
    U_{\mathrm{eff}}
    =
    \begin{pmatrix}
        u_1 & w\\
        w & u_2
    \end{pmatrix}.
\end{equation*}
The quartic-order energy relative to the parent branch is therefore
\begin{equation}
    \Delta f_4(x_1,x_2;T)=\frac12r_1x_1+\frac12r_2x_2+\frac14u_1x_1^2+\frac12wx_1x_2+\frac14u_2x_2^2.\label{eq:two-mode-effective-energy}
\end{equation}
If \(G_i\) denotes the \(i\)-th column of \(G\), then
\begin{align*}
    r_i=A_i(\overline{\by}_\alpha,T),\quad  u_i&=U_{ii}(\overline{\by}_\alpha,T)-\frac12G_i^TH_\alpha^{-1}G_i,\qquad i=1,2
\end{align*}
and
\begin{equation}
    w=U_{12}(\overline{\by}_\alpha,T)-\frac12G_1^TH_\alpha^{-1}G_2.\label{eq:w-def}
\end{equation}
Thus \(u_1\), \(u_2\), and \(w\) contain both their bare contributions and the correction generated by relaxation of the common parent variables. Consequently, parent relaxation can shift the stability boundaries of the single-mode and mixed branches and, through the renormalized mixed coefficient \(w\), modify the balance between cooperative and competitive ordering. For \(x_1,x_2>0\), the mixed term \(wx_1x_2/2\) lowers the energy when \(w<0\) and raises it when \(w>0\); we therefore refer to these cases as cooperative and competitive coupling, respectively.

For later use, define
\begin{equation}
    D:=u_1u_2-w^2.\label{eq:two-mode-D}
\end{equation}

\subsubsection{Single-mode branches and transverse stability: \texorpdfstring{\(\alpha\to\alpha+\xi_1\) or \(\alpha\to\alpha+\xi_2\)}{alpha to alpha + xi1 or alpha + xi2}}

By symmetry under \(1\leftrightarrow2\), it is sufficient to analyze \(\alpha\to\alpha+\xi_1\); the results for \(\alpha+\xi_2\) follow by interchanging the labels.

Assume first that \(u_1>0\). On the single-mode branch \(x_2=0\), stationarity of \eqref{eq:two-mode-effective-energy} with respect to \(x_1\) gives $r_1+u_1x_1=0$. Hence $x_1=-r_1/u_1>0$ with $r_1<0$, and its free energy relative to the parent branch is $\Delta F_1=-r_1^2/4u_1$. Thus \(r_1=0\), \(u_1>0\) gives a continuous single-mode onset. If \(u_1<0\), the quartic truncation is unstable and the transition must be analyzed using the \(2\!-\!4\!-\!6\) reduction of Section~\ref{subsec:246}.

Once the \(\alpha+\xi_1\) branch has formed, we consider the onset of the second mode \(x_2\) while allowing \(x_1\) to relax. From
\[
    \frac{\partial\Delta f_4}{\partial x_1}=\frac12(r_1+u_1x_1+wx_2)=0,
\]
we obtain
\begin{equation*}
    x_1=-\frac{r_1+wx_2}{u_1}=-\frac{r_1}{u_1}-\frac{w}{u_1}x_2.
\end{equation*}
Substituting this relaxed value into
\eqref{eq:two-mode-effective-energy} gives
\begin{equation*}
    \Delta f_4(x_1,x_2)=\Delta F_1+\frac12r_{2|1}x_2+\frac14u_{2|1}x_2^2,
\end{equation*}
where
\begin{equation}
    r_{2|1}:=r_2-\frac{w}{u_1}r_1,\qquad u_{2|1}:=u_2-\frac{w^2}{u_1}=\frac{D}{u_1}.\label{eq:two-mode-transverse-coefficients-r}
\end{equation}
Thus \(r_{2|1}\) is the quadratic curvature of the transverse mode after the first mode has relaxed. The \(\alpha+\xi_1\) branch is locally stable against \(\xi_2\) when $r_{2|1}>0$. A continuous transition to the mixed branch occurs when $r_{2|1}=0$ and $u_{2|1}>0$. If \(u_{2|1}<0\), the transverse onset is first order and must instead be treated by the \(2\!-\!4\!-\!6\) reduction with \(\alpha+\xi_1\) as the new parent branch.

The corresponding formulas for the \(\alpha+\xi_2\) branch follow from the exchange \(1\leftrightarrow2\).

\subsubsection{Mixed branch: \texorpdfstring{\(\alpha\to\alpha+\xi_1+\xi_2\)}{alpha to alpha + xi1 + xi2}}

For an interior mixed state \(x_1>0\), \(x_2>0\), stationarity of \eqref{eq:two-mode-effective-energy} gives
\begin{equation*}
    \begin{pmatrix}
        u_1 & w\\
        w & u_2
    \end{pmatrix}
    \begin{pmatrix}
        x_1\\x_2
    \end{pmatrix}
    =-
    \begin{pmatrix}
        r_1\\r_2
    \end{pmatrix}.
\end{equation*}
For \(D\neq0\), inversion gives
\begin{equation}
    x_1=\frac{wr_2-u_2r_1}{D},\qquad x_2=\frac{wr_1-u_1r_2}{D}.\label{eq:two-mode-mixed-branch}
\end{equation}
At quartic order, this interior stationary state is a local minimum when
\begin{equation}
    u_1>0,\qquad u_2>0,\qquad D>0,\qquad x_1>0,\qquad x_2>0.\label{eq:mix-branch-local-stable}
\end{equation}

Using \(U_{\mathrm{eff}}\bm x=-\bm r\) on the mixed branch, its free energy can be written as $\Delta F_{12}=\bm r\cdot\bm x/4$, and hence
\begin{equation*}
    \Delta F_{12}=-\frac{u_2r_1^2+u_1r_2^2-2wr_1r_2}{4D}.
\end{equation*}

The mixed branch connects continuously to the single-mode \(\alpha+\xi_1\) branch when the second mode first becomes unstable. Indeed, if
\[
    r_{2|1}=r_2-\frac{w}{u_1}r_1=0,
\]
then \eqref{eq:two-mode-mixed-branch} gives $x_2=0$ and $x_1=-r_1/u_1$, which is precisely the single-mode solution on the \(\alpha+\xi_1\) branch. The corresponding result for \(\alpha+\xi_2\) follows by exchanging \(1\leftrightarrow2\).

Whenever the quartic theory is stable, the locally preferred state is obtained by comparing $0$, $\Delta F_1$, $\Delta F_2$ and $\Delta F_{12}$. This comparison is local to the reduced two-mode theory; the globally realized phase is ultimately determined from the fully relaxed branch free energies.

\paragraph{Direct continuous transition.}
A continuous transition directly from the parent branch into the mixed state requires both modes to become critical at the same temperature:
\begin{equation}
    r_1(T_c)=r_2(T_c)=0.\label{eq:continuous-direct-mixed}
\end{equation}
The quartic form must be positive definite,
\begin{equation*}
    u_1(T_c)>0,\qquad u_2(T_c)>0,\qquad D(T_c)>0,
\end{equation*}
Upon cooling through \(T_c\), the mixed branch must also be physically admissible, namely \(x_1>0\) and \(x_2>0\) in \eqref{eq:two-mode-mixed-branch}. Thus the conditions \(r_1(T_c)=r_2(T_c)=0\) are necessary but not sufficient: the cooling trajectory must enter the region where both ordering modes are nonzero. Since these are two simultaneous conditions, such a direct continuous transition is nongeneric when temperature is the only control parameter and generally requires an additional tuning parameter.

\paragraph{Direct first-order transition.}
A direct first-order transition need not satisfy \eqref{eq:continuous-direct-mixed}. To analyze a finite-amplitude mixed
onset, parameterize the positive quadrant by
\begin{equation*}
    x_1=R^2\cos^2\varphi,\qquad x_2=R^2\sin^2\varphi,\qquad R\geq0,\quad0<\varphi<\frac{\pi}{2}.
\end{equation*}
Define
\[
    q(\varphi):=
    \begin{pmatrix}
        \cos^2\varphi\\
        \sin^2\varphi
    \end{pmatrix},
    \qquad \bm x=R^2q(\varphi).
\]
Along a fixed direction \(\varphi\), the two-mode problem \eqref{eq:universal-energy-2} reduces to the one-amplitude expansion
\begin{equation}
    \Delta f(R,\varphi;T)=\frac12r(\varphi;T)R^2+\frac14u(\varphi;T)R^4+\frac16v(\varphi;T)R^6+O(R^8),\label{eq:phi-energy}
\end{equation}
with
\begin{equation*}
    r(\varphi;T)=\bm r\cdot q(\varphi)=r_1\cos^2\varphi+r_2\sin^2\varphi,
\end{equation*}
and
\begin{equation*}
    u(\varphi;T)=q(\varphi)^TU_{\mathrm{eff}}q(\varphi)=u_1\cos^4\varphi+u_2\sin^4\varphi+2w\cos^2\varphi\sin^2\varphi.
\end{equation*}

The corresponding parent-variable relaxation is
\[
    \by-\overline{\by}_\alpha=\by_\varphi R^2+O(R^4),\qquad\by_\varphi:=-\frac12H_\alpha^{-1}G_\varphi,
\]
where $G_\varphi:=Gq(\varphi)$. Thus the directional problem has exactly the same form as the one-mode \(2\!-\!4\!-\!6\) reduction of Section~\ref{subsec:246}. To write its sixth-order coefficient, define
\[
    M_\varphi:=\cos^2\varphi\,D_{\by}^2A_1+\sin^2\varphi\,D_{\by}^2A_2,\quad \bm h_\varphi:=\nabla_{\by}\left(q(\varphi)^TUq(\varphi)\right).
\]
Let the bare directional sixth-order coefficient \(C_\varphi\) be defined by
\begin{equation*}
    f_6\bigl(R^2q(\varphi);\overline{\by}_\alpha,T\bigr)=\frac16C_\varphi(T)R^6.
\end{equation*}
Applying the one-mode result \eqref{eq:246-v} to this fixed direction gives
\begin{equation*}
    v(\varphi;T)=\,C_\varphi+\mathcal T_\alpha[\by_\varphi,\by_\varphi,\by_\varphi]+\frac32\by_\varphi^TM_\varphi\by_\varphi+\frac32\bm h_\varphi\cdot\by_\varphi.
\end{equation*}

If an interior direction \(\varphi_c\) satisfies $u(\varphi_c;T_{\rm coex})<0$ and $v(\varphi_c;T_{\rm coex})>0$, then the \(2\!-\!4\!-\!6\) coexistence conditions give
\begin{equation}
    R_c^2=-\frac{3u(\varphi_c;T_{\rm coex})}{4v(\varphi_c;T_{\rm coex})},\qquad r(\varphi_c;T_{\rm coex})=\frac{3u^2(\varphi_c;T_{\rm coex})}{16v(\varphi_c;T_{\rm coex})}.\label{eq:phi-c}
\end{equation}
Since \(\varphi\) is itself free to relax, the selected mixed direction must additionally satisfy
\begin{equation}
    \partial_\varphi\Delta f(R_c,\varphi_c;T_{\rm coex})=0,\qquad\partial_{\varphi\varphi}\Delta f(R_c,\varphi_c;T_{\rm coex})>0.\label{eq:phi-angular-stability}
\end{equation}
Finally, let \(F_{\alpha+\xi_1+\xi_2}(T)\) denote the fully relaxed free energy of the finite-amplitude mixed branch. At coexistence, this branch must have the same free energy as the parent phase and lie below every other locally stable competing branch:
\begin{equation}
\begin{aligned}
F_{\alpha+\xi_1+\xi_2}(T_{\rm coex})
  &= F_\alpha(T_{\rm coex}),\\
F_{\alpha+\xi_1+\xi_2}(T_{\rm coex})
  &< F_\beta(T_{\rm coex})
     \quad\bigl(\forall\,\beta\notin
     \{\alpha,\alpha+\xi_1+\xi_2\}\bigr).
\end{aligned}
\label{eq:direct-first-order-global-selection}
\end{equation}
Thus a direct first-order mixed transition is selected at finite amplitude and does not require simultaneous linear instability of the two modes.

\paragraph{Local competition regimes.}

Assuming \(u_1>0\) and \(u_2>0\), the local competition is classified by the quartic determinant \(D\) defined in \eqref{eq:two-mode-D} and by the sign of \(w\) defined in \eqref{eq:w-def}. For \(D>0\), the quartic theory can support a locally stable mixed branch whenever \eqref{eq:mix-branch-local-stable} is satisfied. For \(D<0\), no locally stable interior mixed branch exists at quartic order. The resulting cooperative and competitive regimes are summarized in Table~\ref{tab:two-mode-coupling-regimes}.

\begin{table}[H]
\centering
\small
\renewcommand{\arraystretch}{1.35}
\setlength{\tabcolsep}{3pt}
\caption{Local two-mode regimes for \(u_1,u_2>0\).}
\label{tab:two-mode-coupling-regimes}

\begin{tabular}{@{}
>{\raggedright\arraybackslash}p{1.6cm}
>{\raggedright\arraybackslash}p{2.7cm}
>{\raggedright\arraybackslash}p{2.2cm}
>{\raggedright\arraybackslash}p{5.1cm}@{}}
\toprule
Mixed-branch condition & Coupling range & Interaction & Local consequence \\
\midrule

\multirow{2}{*}{\(D>0\)} & \(-\sqrt{u_1u_2}<w<0\) & Cooperative & Stable mixed branch possible; the first mode promotes the second.\\
\cline{2-4}

& \(0<w<\sqrt{u_1u_2}\) & Weakly competitive & Stable mixed branch possible; the first mode delays the second.\\
\midrule

\multirow{2}{*}{\(D<0\)} & \(w<-\sqrt{u_1u_2}\) & Strongly cooperative & Quartic instability; sixth-order terms may stabilize direct mixed ordering.\\
\cline{2-4}

& \(w>\sqrt{u_1u_2}\) & Strongly competitive & Mixed branch unstable; the two single-mode branches compete. \\

\bottomrule
\end{tabular}
\end{table}

The boundary cases \(w=0\) and \(D=0\) correspond respectively to uncoupled and marginal quartic theories. In the strongly cooperative \(D<0\) regime, the directional sixth-order reduction \eqref{eq:phi-energy}--\eqref{eq:direct-first-order-global-selection} determines whether a finite-amplitude mixed state is stabilized. In the strongly competitive regime, the competing single-mode branches must instead be compared through their fully relaxed free energies. These local regimes provide the input to the pathway-selection procedure below.

\subsubsection{Selection of two-mode ordering pathways}

A continuous onset is determined by loss of local stability, whereas a first-order onset is determined by finite-amplitude coexistence. Consequently, the first mode to order is not necessarily the one whose quadratic coefficient vanishes first.

For \(i=1,2\), let \(T_i^{\mathrm{cand}}\) denote the highest physically admissible onset temperature of the single-mode branch \(\alpha+\xi_i\). The continuous and first-order candidates are determined respectively by
\begin{equation}
\begin{cases}
    r_i(T_i^{\mathrm{cand}})=0,\qquad u_i(T_i^{\mathrm{cand}})>0, & \text{continuous onset},\\[2mm]
    \displaystyle r_i(T_i^{\mathrm{cand}})=\frac{3u_i^2(T_i^{\mathrm{cand}})}{16v_i(T_i^{\mathrm{cand}})},\qquad u_i(T_i^{\mathrm{cand}})<0,\quad v_i(T_i^{\mathrm{cand}})>0, & \text{first-order onset}.
\end{cases}
\label{eq:two-mode-single-candidate-temperature}
\end{equation}
The corresponding onset value of \(x_i=\xi_i^2\) is
\begin{equation*}
    x_i(T_i^{\mathrm{cand}})=\begin{cases}
    0^+, & \text{continuous onset},\\[2mm]
    \displaystyle-\frac{3u_i(T_i^{\mathrm{cand}})}{4v_i(T_i^{\mathrm{cand}})},& \text{first-order onset}.
\end{cases}
\end{equation*}
The tricritical boundary is given by $r_i=u_i=0$ with $v_i>0$.

To include direct mixed ordering, let \(T_{12}^{\rm cand}\) denote the highest admissible onset temperature of the mixed branch. Continuous and first-order mixed onsets are determined by the corresponding criteria derived above; if no admissible mixed onset exists, we set \(T_{12}^{\rm cand}=-\infty\).

The first transition encountered upon cooling is determined by the largest of \(T_1^{\rm cand}\), \(T_2^{\rm cand}\), and \(T_{12}^{\rm cand}\). If a single-mode branch \(\alpha+\xi_i\) appears first, it becomes the new parent branch, and the subsequent onset of the second mode is determined by the relaxed coefficients \(r_{j|i}\), \(u_{j|i}\), and \(v_{j|i}\). Table~\ref{tab:two-mode-sequence-selection} summarizes these possibilities.

\begin{table}[H]
\centering
\footnotesize
\setlength{\tabcolsep}{2pt}
\renewcommand{\arraystretch}{1.30}
\caption{Two-mode ordering pathways and transition criteria.}
\label{tab:two-mode-sequence-selection}

\begin{tabular}{@{}
>{\raggedright\arraybackslash}p{2.3cm}
>{\raggedright\arraybackslash}p{2.5cm}
>{\raggedright\arraybackslash}p{2.2cm}
>{\raggedright\arraybackslash}p{4.8cm}@{}}
\toprule
Pathway & Pathway criterion & Phase transition type & Transition criterion \\
\midrule

\multirow{2}{*}{\makecell[l]{Single-mode\\ordering:\\
\(\alpha\to\alpha+\xi_i\)\\
\(\to\alpha+\xi_1+\xi_2\)}}
&
\multirow{2}{*}{\makecell[l]{
\(T_i^{\mathrm{cand}}>T_j^{\mathrm{cand}}\)\\
and \\
\(T_i^{\mathrm{cand}}>T_{12}^{\mathrm{cand}}\)}}
&
\makecell[l]{First onset:\\
\(\alpha\to\alpha+\xi_i\)}
&
Continuous: \(r_i=0,\ u_i>0\).\par
First order: \(u_i<0,\ v_i>0,\ r_i=3u_i^2/(16v_i)\).
\\
\cline{3-4}

& &
\makecell[l]{Subsequent onset:\\
\(\alpha+\xi_i\to\)\\
\(\alpha+\xi_1+\xi_2\)}
&
Coefficients are evaluated on the fully relaxed
\(\alpha+\xi_i\) branch.\par
Continuous: \(r_{j|i}=0,\ u_{j|i}>0\).\par
First order: \(u_{j|i}<0,\ v_{j|i}>0,\
r_{j|i}=3u_{j|i}^2/(16v_{j|i})\).
\\
\midrule

\multirow{2}{*}{\makecell[l]{Direct mixed\\ordering:\\
\(\alpha\to\)\\
\(\alpha+\xi_1+\xi_2\)}}
&
\multirow{2}{*}{\makecell[l]{
\(T_{12}^{\mathrm{cand}}>T_1^{\mathrm{cand}}\)\\
and\\
\(T_{12}^{\mathrm{cand}}>T_2^{\mathrm{cand}}\)}}
&
Direct continuous
&
\(r_1(T_c)=r_2(T_c)=0\), \(u_1,u_2>0\), \(D>0\),
with the cooling trajectory entering \(x_1,x_2>0\).
\\
\cline{3-4}

& &
Direct first order
&
For some \(\varphi_c\in(0,\pi/2)\),
Eqs.~\eqref{eq:phi-c},
\eqref{eq:phi-angular-stability}, and
\eqref{eq:direct-first-order-global-selection}.
\\
\bottomrule
\end{tabular}
\end{table}
Here \(r_{j|i}\), \(u_{j|i}\), and \(v_{j|i}\) are the corresponding coefficients evaluated on the relaxed \(\alpha+\xi_i\) branch. Near the original parent branch, \(r_{j|i}\) and \(u_{j|i}\) reduce to the transverse coefficients in \eqref{eq:two-mode-transverse-coefficients-r}; the equilibrium sequence is then selected by the fully relaxed branch free energies.

\subsection{Modular phase selection}\label{subsec:branch-adaptive-reduction}
The conditions in Table~\ref{tab:two-mode-sequence-selection} identify locally admissible pathways but do not by themselves determine the equilibrium sequence. For each current parent branch \(\alpha\), the modular phase-selection procedure is:

\begin{enumerate}
    \item Solve \eqref{eq:parent-branch} and retain only locally stable parent branches.
    \item Identify the one-mode or two-mode ordering coordinates of every candidate daughter branch.
    \item Relax the parent variables using \eqref{eq:general-parent-relaxation} and classify the local transitions using Section~\ref{subsec:one-mode} and \ref{subsec:two-mode}.
    \item Compare the fully relaxed free energies of all locally stable branches. At each temperature, select
    \begin{equation}
    \alpha_{\mathrm{eq}}(T)\in\operatorname*{arg\,min}_{\beta\in\mathcal B_{\mathrm{loc}}(T)} F_\beta(T),\label{eq:branch-adaptive-global-selection}
    \end{equation}
    where \(\mathcal B_{\mathrm{loc}}(T)\) is the set of locally stable, physically admissible, and fully relaxed branches. Promote the selected branch \(\alpha_{\mathrm{eq}}(T)\) to the new parent branch and repeat the procedure.
\end{enumerate}

The local criteria generate candidate daughter branches, while \eqref{eq:branch-adaptive-global-selection} selects the realized branch. Fig~\ref{fig:modular} summarizes this modular construction, in which each selected daughter branch becomes the parent for the next local transition.

\section{Application to ferroelectric liquid-crystal phase transitions}\label{sec:liquid-crystal-application}

The modular framework of Section~\ref{sec:universal-landau-reductions} can be applied successively to liquid-crystal phase sequences. For example, $I\to N\to N_F\to\SmA_F\to\SmC_F$ is assembled from the local transitions $I\to N$, $N\to N_F$, $N_F\to\SmA_F$ and $\SmA_F\to\SmC_F$. The corresponding modules are developed below, with their coefficients evaluated on the relevant relaxed parent branch. Other admissible sequences are assembled analogously, so the subsections below are organized by individual transitions rather than by complete cooling pathways.

\subsection{Modelling of liquid crystal in Landau framework}\label{sec:general-model}

We now give a concrete liquid-crystal realization of the modular Landau framework developed in Section~\ref{sec:universal-landau-reductions}. The phases considered below involve apolar orientational order, polar order, smectic layering, and director--layer tilt. We represent these orders by a symmetric traceless nematic tensor \(Q\), a polarization vector \(P\), and a complex smectic order parameter \(\psi=\rho e^{i\varphi}\), with the tilt encoded by the relative orientation between the nematic director and the smectic wave vector. The \(Q\)-tensor description follows Landau--de Gennes theory~\cite{deGennesProst1993,deGennes1971}; alternative descriptions of nematic order include director theories \cite{Oseen1933,Frank1958}, variable-degree models \cite{Ericksen1991}, molecular orientational distribution functions \cite{Onsager1949}, scalar uniaxial reductions \cite{Gramsbergen1986}, and higher-rank tensor descriptions \cite{LubenskyRadzihovsky2002}. Polar order may likewise be represented by a macroscopic polarization vector \cite{Kats2021,ZouAya2024} or, under stronger orientational locking, by a scalar polar amplitude \cite{Etxebarria2022,Sebastian2023}. For smectic order, the complex density-wave amplitude adopted here follows the classical de Gennes--Chen--Lubensky construction \cite{ChenLubensky1976,Biscari2007,deGennes1972}, while alternative formulations use real-density fields \cite{ShiHan2025,XiaHan2024,Pevnyi2014}, layer-displacement fields \cite{deGennesProst1993}, or tensorial smectic variables \cite{Paget2022}. \textit{The \(Q+P+\psi\) formulation used here is one possible description of the relevant liquid-crystal orders, and other choices of order parameters are also available.}

With these order parameters, we write the free energy as
\begin{equation*}
    \mathcal F[Q,P,\psi]=\int_\Omega\left(f_{\mathrm{bulk}}+f_{\mathrm{el}}+f_{\mathrm{cpl}}\right)\,\mathrm{d}x .
\end{equation*}
Here \(f_{\mathrm{bulk}}\) describes the formation of nematic, polar, and smectic order, \(f_{\mathrm{el}}\) contains the spatial and wavelength-selection contributions, and \(f_{\mathrm{cpl}}\) describes the interactions among these orders.

The bulk energy is written as $f_{\mathrm{bulk}}=f_Q+f_P+f_\psi$,  with Landau expansions for the three order parameters introduced above. For the nematic tensor, the cubic rotational invariant \(\operatorname{tr}(Q^3)\) is allowed, and we adopt the Landau--de Gennes \(2\!-\!3\!-\!4\) form \cite{deGennesProst1993,deGennes1971},
\begin{equation}
    f_Q(Q;T)=\frac{A_Q(T)}{2}\operatorname{tr}(Q^2)-\frac{B_Q}{3}\operatorname{tr}(Q^3) +\frac{C_Q}{4}\bigl(\operatorname{tr}(Q^2)\bigr)^2 .\label{eq:bulk-Q}
\end{equation}
This is the standard lowest-order stable tensor expansion for the isotropic--nematic transition and gives the generic first-order transition of Landau--de Gennes theory.

The polar and smectic order parameters have the symmetries \(P\mapsto-P\) and \(\psi\mapsto e^{i\alpha}\psi\), respectively, so their local free energies contain only even powers of the corresponding amplitudes. We take
\begin{align}
    f_P(P;T)&=\frac{A_P(T)}{2}|P|^2+\frac{B_P}{4}|P|^4+\frac{C_P}{6}|P|^6,\label{eq:bulk-P}\\
    f_\psi(\psi;T)&=\frac{A_\psi(T)}{2}|\psi|^2+\frac{B_\psi}{4}|\psi|^4+\frac{C_\psi}{6}|\psi|^6 .\label{eq:bulk-psi}
\end{align}
The even \(2\!-\!4\!-\!6\) form is the minimal expansion that can describe continuous, tricritical, and first-order transitions while remaining bounded when the quartic coefficient becomes negative \cite{Landau1937,Devonshire1949,Cowley1980}. Complex-amplitude Landau expansions of this type are standard in continuum descriptions of smectic ordering \cite{deGennes1972,ChenLubensky1976}.

The quadratic coefficients are taken as $A_i(T)=a_i(T-T_i^*),$ for $i\in\{Q,P,\psi\}$ and $a_i>0,$ while the higher-order coefficients are treated as temperature independent over the range considered. We assume \(B_Q,C_Q,C_P,C_\psi>0\), whereas \(B_P\) and \(B_\psi\) may have either sign.

For elastic and wavelength-selection terms, standard gradient terms for \(Q\) and \(P\) may be included to describe nematic and polar textures \cite{deGennesProst1993,Frank1958,Longa1987,MottramNewton2014,Shamid2013,Mertelj2018,PaikSelinger2025}. Since they do not enter the homogeneous branch analysis below, we do not write them explicitly here.

For the smectic order, we retain the isotropic Brazovskii--Swift--Hohenberg wavelength-selection term \cite{Brazovskii1975,SwiftHohenberg1977}, $f_{\mathrm{el}}=K_\Psi\left|(\nabla^2+q_0^2)\psi\right|^2/2$. For the plane wave \(\psi=\rho e^{iq\mathbf m\cdot x}\), \(|\mathbf m|=1\), this reduces to
\begin{equation}\label{eq:wavelength-penalty-reduction}
    f_{\mathrm{el}}=\frac{K_\psi}{2}(q^2-q_0^2)^2\rho^2,
\end{equation}
and therefore selects the preferred bare wave number \(q_0\). Alternative smectic gradient constructions include the director-covariant de Gennes form, the longitudinal--transverse Chen--Lubensky model, and tensor--density formulations \cite{deGennes1972,Biscari2007,ChenLubensky1976,XiaHan2024,ShiHan2025}.

To couple the nematic, polar, and smectic orders, we retain the lowest-order amplitude and orientational interactions needed for the phase sequences considered below:
\begin{align}
    f_{\mathrm{cpl}}=&-\lambda P_iQ_{ij}P_j-\chi_P|\psi|^2|P|^2-g_1Q_{ij}(\partial_i\psi)^*(\partial_j\psi)\nonumber\\
    &\quad+g_PP_iP_j(\partial_i\psi)^*(\partial_j\psi)+\frac{g_2}{2}\left|Q_{ij}(\partial_i\psi)^*(\partial_j\psi)\right|^2 .\label{eq:coupling-terms}
\end{align}
The \(P\)--\(Q\) coupling proportional to \(\lambda\) favors polarization along the principal nematic axis for \(\lambda>0\), considered in isolation, while the biquadratic term proportional to \(\chi_P\) couples the polar and smectic amplitudes \cite{ChaikinLubensky1995,Longa1987,LongaTrebin1990,PajakLonga2018,Nikolova2025}. The longitudinal constraint \(P=pn\) adopted below is imposed as part of the reduced model, rather than inferred from the sign of \(\lambda\) alone.

The remaining terms control the relative orientation of the nematic director, polarization, and smectic wave vector. Nematic--smectic gradient couplings of this type are standard in continuum theories \cite{deGennes1972,Biscari2007,ChenLubensky1976}. In the present model, the \(g_1>0\) term favors nematic--layer alignment, \(g_P\) introduces the polar contribution to this orientational coupling. For \(g_2>0\), the additional \(g_2\) term allows the angular energy to develop a finite-tilt minimum \cite{Izzo2020,ChenLubensky1976,ShiHan2025,XiaHan2024}.

For the homogeneous phases considered below, we adopt the constrained uniaxial, longitudinal, single-wave-vector ansatz
\begin{equation}
    Q=S\left(n\otimes n-\frac{\mathbf I}{3}\right),\qquad P=pn,\qquad\psi=\rho e^{iqm\cdot x},\qquad\cos\theta=n\cdot m .\label{eq:uniaxial-plane-wave-ansatz}
\end{equation}
Here \(S\) and \(p\) are the scalar nematic and polar order parameters, \(\rho=|\psi|\) is the smectic order-parameter amplitude, \(q\) is the smectic wave number, and \(\theta\) is the director--layer angle. In particular, the constraint \(P=pn\) is imposed as part of the reduced model, so the polarization direction is not varied independently of \(n\). Accordingly, stability and phase-selection results below refer to the retained degrees of freedom within this constrained homogeneous ansatz.

Substituting \eqref{eq:uniaxial-plane-wave-ansatz} into the bulk energies \eqref{eq:bulk-Q}--\eqref{eq:bulk-psi}, the wavelength-selection term \eqref{eq:wavelength-penalty-reduction}, and the couplings \eqref{eq:coupling-terms}, we obtain
\begin{align}
  f(S,p,\rho,q,\theta;T)=&\frac{A_Q}{3}S^2-\frac{2B_Q}{27}S^3+\frac{C_Q}{9}S^4+\frac{A_P}{2}p^2+\frac{B_P}{4}p^4+\frac{C_P}{6}p^6-\frac{2\lambda}{3}Sp^2 \nonumber\\
  &+\frac{A_\psi}{2}\rho^2+\frac{B_\psi}{4}\rho^4+\frac{C_\psi}{6}\rho^6-\chi_Pp^2\rho^2+\frac{K_\psi}{2}(q^2-q_0^2)^2\rho^2 \nonumber\\
  &+q^2\rho^2\Bigl[-g_1S\bigl(\cos^2\theta-\frac13\bigr)+g_Pp^2\cos^2\theta+\tfrac{g_2}{2}q^2S^2\rho^2\bigl(\cos^2\theta-\frac13\bigr)^2\Bigr].\label{eq:plane-wave-energy}
\end{align}

For the homogeneous thermodynamic analysis below, the smectic modulation wave number is fixed at its preferred reference value \(q=q_0\), so that the wavelength-selection term in \eqref{eq:plane-wave-energy} vanishes identically. The spatial coupling coefficients then enter the reduced free energy only through the combinations
\begin{equation*}
    g_1^{\mathrm{eff}}:=g_1q_0^2,\qquad g_P^{\mathrm{eff}}:=g_Pq_0^2,\qquad g_2^{\mathrm{eff}}:=g_2q_0^4 .\end{equation*}
Here the coefficients on the right-hand side are those appearing in the spatial free energy. For notational simplicity, throughout the reduced thermodynamic analysis we reuse the symbols $(g_1,g_P,g_2)\equiv(g_1^{\mathrm{eff}},g_P^{\mathrm{eff}},g_2^{\mathrm{eff}})$. Thus \(q_0\) does not appear as an independent parameter in the reduced thermodynamic analysis. All subsequent parameter variations refer to these effective fixed-wave-number couplings.

With this convention, the homogeneous thermodynamic free energy becomes
\begin{align}
	f(S,p,\rho,\theta;T)=&\frac{A_Q}{3}S^2-\frac{2B_Q}{27}S^3+\frac{C_Q}{9}S^4+\frac{A_P}{2}p^2+\frac{B_P}{4}p^4 \nonumber\\
	&+\frac{C_P}{6}p^6-\frac{2\lambda}{3}Sp^2+\frac{A_\psi}{2}\rho^2+\frac{B_\psi}{4}\rho^4+\frac{C_\psi}{6}\rho^6-\chi_Pp^2\rho^2 \nonumber\\
	&+\rho^2\Bigl[-g_1S\bigl(\cos^2\theta-\frac13\bigr)+g_Pp^2\cos^2\theta+\frac{g_2}{2}S^2\rho^2 \bigl(\cos^2\theta-\frac13\bigr)^2\Bigr].
\label{eq:reduced-energy}
\end{align}

For \(S\rho>0\), define \(z=\cos^2\theta\in[0,1]\). The angular part of \eqref{eq:reduced-energy} is
\[
    f_{\mathrm{ang}}(z)=-g_1S\rho^2\left(z-\frac13\right)+g_Pp^2\rho^2z+\frac{g_2}{2}S^2\rho^4\left(z-\frac13\right)^2 .
\]
For \(g_2>0\), its constrained minimizer is
\[
    z_*=\Pi_{[0,1]}\left(\frac13+\frac{g_1S-g_Pp^2}{g_2S^2\rho^2}\right).
\]
where \(\Pi_{[0,1]}\) denotes projection onto the interval \([0,1]\). In the parameter regimes considered below, \(z_*=1\) and \(0<z_*<1\) correspond respectively to \(\SmA\)-type and \(\SmC\)-type layer orientations. The phases considered in the following analysis are summarized in Table~\ref{tab:phase-identification}.

\begin{table}[H]
\centering
\small
\renewcommand{\arraystretch}{1.2}
\caption{Homogeneous phases in the uniaxial plane-wave reduction.}
\label{tab:phase-identification}

\begin{tabular*}{\textwidth}{@{\extracolsep{\fill}}ccccc@{}}
\toprule
Phase & \(S\) & \(|p|\) & \(\rho\) & \(z=\cos^2\theta\) \\
\midrule
\(I\)       & \(0\)  & \(0\)  & \(0\)  & -- \\
\(N\)       & \(>0\) & \(0\)  & \(0\)  & -- \\
\(\NF\)     & \(>0\) & \(>0\) & \(0\)  & -- \\
\(\SmA\)    & \(>0\) & \(0\)  & \(>0\) & \(1\) \\
\(\SmAF\)   & \(>0\) & \(>0\) & \(>0\) & \(1\) \\
\(\SmC\)    & \(>0\) & \(0\)  & \(>0\) & \(0<z<1\) \\
\(\SmCF\)   & \(>0\) & \(>0\) & \(>0\) & \(0<z<1\) \\
\bottomrule
\end{tabular*}
\end{table}

\subsection{Applications of the modular phase transition analysis to liquid crystals}
\label{sec:N-to-SmAF}

This section applies the modular reductions developed in Section~\ref{sec:universal-landau-reductions} to liquid-crystal phase sequences. The isotropic--nematic transition is described by the \(2\!-\!3\!-\!4\) model, while polar, smectic, and tilt onsets follow the one-mode \(2\!-\!4\!-\!6\) reduction. The two-mode reductions, combined with modular phase selection, then determine cooperative ordering, competing pathways, and the equilibrium sequence selected by the fully relaxed branch free energies.

\subsubsection{\textbf{Application of the} \texorpdfstring{\(2\!-\!3\!-\!4\)}{2-3-4} \textbf{model}: \texorpdfstring{\boldmath \(I\to N\)}{I to N}}
\label{par:IN-transition-example}

Substituting the uniaxial ansatz \eqref{eq:uniaxial-plane-wave-ansatz} into the nematic bulk energy \eqref{eq:bulk-Q} gives 
\begin{equation*}
    f_{IN}(S;T)=\frac{A_Q(T)}{3}S^2-\frac{2B_Q}{27}S^3+\frac{C_Q}{9}S^4 .
\end{equation*}
Comparison with the \(2\!-\!3\!-\!4\) normal form \eqref{eq:234-reduced-energy} identifies $r_S=2A_Q/3$, $b_S=2B_Q/9$ and $u_S=4C_Q/9$. The locally stable prolate nematic branch is therefore
\begin{equation}\label{eq:I-N-stable-branch}
    S_N(T)=\frac{B_Q+\sqrt{B_Q^2-24C_QA_Q(T)}}{4C_Q}.
\end{equation}

For subsequent transitions out of the nematic phase, the required
parent-branch derivatives are
\begin{subequations}\label{eq:IN-parent-derivatives}
\begin{align}
    H_N(T)&:=f_{IN}''(S_N(T);T)=\frac{2S_N(T)}{9}\left(4C_QS_N(T)-B_Q\right)>0,\label{eq:I-N-Hdef}\\
    J_N(T)&:=f_{IN}'''(S_N(T);T)=-\frac{4B_Q}{9}+\frac{8C_Q}{3}S_N(T).\label{eq:I-N-Jdef}
\end{align}
\end{subequations}
These quantities are the scalar counterparts of the parent Hessian and third derivative in Section~\ref{subsec:parent-branch-relaxation}.

Applying the general spinodal and coexistence conditions of Section~\ref{subsec:234} gives
\begin{equation*}
    T_{\mathrm{sp}}^{IN}=T_Q^*+\frac{B_Q^2}{24C_Qa_Q},\quad  S_{IN}=\frac{B_Q}{3C_Q},\quad T_{\mathrm{coex}}^{IN}=T_{IN}=T_Q^*+\frac{B_Q^2}{27C_Qa_Q},\quad T_c^{IN}=T_Q^*.
\end{equation*}
Consequently, $T_{\mathrm{sp}}^{IN}>T_{\rm coex}^{IN}>T_c^{IN}=T_Q^*$, giving the characteristic hierarchy of the  \(2\!-\!3\!-\!4\) reduction.

\subsubsection{\textbf{Application of the} \texorpdfstring{\(2\!-\!4\!-\!6\)}{2-4-6} \textbf{model}: \texorpdfstring{\boldmath \(N\to N_F/\SmA\to \SmA_F\), \textbf{and} \(\SmA_F\to\SmC_F\)}{N to NF/SmA, NF/SmA to SmAF, and SmAF to SmCF}}
\label{sec:one-mode-applications}

We apply the one-mode \(2\!-\!4\!-\!6\) reduction to polar, smectic, and tilt instabilities. For each transition, we first identify the relevant coefficients before parent-branch relaxation and then obtain the effective coefficients \(r\), \(u\), and \(v\) using \eqref{eq:246-ru} and \eqref{eq:246-v}.

\paragraph{\texorpdfstring{\boldmath \(N\to N_F\) and \(N\to\SmA\)}{N to NF and N to SmA}}
\label{subsec:one-mode-near-N}
\hfill\break

For \(N\to N_F\), with ordering coordinate \(p\), the coefficients before relaxation of the nematic parent branch are

\[
    A_{F|N}(S;T)=A_P(T)-\frac{4\lambda}{3}S,\qquad B_{F|N}=B_P,\qquad C_{F|N}=C_P,
\]
with $g_{F|N}:=\partial_S A_{F|N}=-4\lambda/3$. 

For \(N\to\SmA\), with ordering coordinate \(\rho\), the corresponding coefficients are
\[
    A_{A|N}(S;T)=A_\psi(T)-\frac{4g_1}{3}S,\qquad B_{A|N}(S)=B_\psi+\frac{8g_2}{9}S^2,\qquad C_{A|N}=C_\psi,
\]
with $g_{A|N}:=\partial_S A_{A|N}=-4g_1/3$.

Using \(S_N(T)\), \(H_N(T)\), and \(J_N(T)\) from \eqref{eq:I-N-stable-branch}, \eqref{eq:I-N-Hdef} and \eqref{eq:I-N-Jdef}, the polar and smectic onsets are written as
\begin{equation*}
    \Delta f_j^{\mathrm{eff}}(\xi_j;T)=\frac12 r_j(T)\xi_j^2+\frac14 u_j(T)\xi_j^4+\frac16 v_j(T)\xi_j^6+O(\xi_j^8),\qquad(\xi_F,\xi_A)=(p,\rho).
\end{equation*}
For \(N\to N_F\), substitution into \eqref{eq:246-ru} and \eqref{eq:246-v} gives
\begin{align}
    r_F=A_P(T)-\frac{4\lambda}{3}S_N,\quad u_F=B_P-\frac{8\lambda^2}{9H_N},\quad v_F=C_P+\frac{8J_N\lambda^3}{27H_N^3}.\label{eq:near-N-polar-coefficients}
\end{align}
For \(N\to\SmA\), the corresponding coefficients are
\begin{align}
    &r_A=A_\psi(T)-\frac{4g_1}{3}S_N,\quad u_A=B_\psi+\frac{8g_2}{9}S_N^2-\frac{8g_1^2}{9H_N},\nonumber\\
    &v_A=C_\psi+\frac{8J_Ng_1^3}{27H_N^3}+\frac{16g_1g_2S_N}{9H_N}.\label{eq:near-N-smectic-coefficients}
\end{align}

\paragraph{\texorpdfstring{\boldmath \(N_F\to\SmA_F\) and \(\SmA\to\SmA_F\)}{NF to SmAF and SmA to SmAF}}
\label{sec:away-nematic}
\hfill\break

Define the restricted parent-phase free energies 
\[
    f_{N_F}(S,p;T):=f(S,p,0,0;T),\qquad f_{\SmA}(S,\rho;T):=f(S,0,\rho,0;T),
\]
and let $\mathbf y_F=(S_F,p_F)$, $\mathbf y_A=(S_A,\rho_A)$ denote their relaxed equilibrium branches. They satisfy
\begin{align*}
    \nabla_{\mathbf y_F}f_{N_F}(\mathbf y_F;T)=0,\quad H_F:=D_{\mathbf y_F}^2f_{N_F}(\mathbf y_F;T)\succ0;\\
    \nabla_{\mathbf y_A}f_{\SmA}(\mathbf y_A;T)=0,\quad H_A:=D_{\mathbf y_A}^2f_{\SmA}(\mathbf y_A;T)\succ0.
\end{align*}

For the polar-first pathway \(N_F\to\SmA_F\), with ordering coordinate
\(\rho\), the coefficients before parent-branch relaxation are
\[
    A_{A|F}(S,p;T)=A_\psi(T)-\frac{4g_1}{3}S+2(g_P-\chi_P)p^2,\quad B_{A|F}(S,p)=B_\psi+\frac{8g_2}{9}S^2,\quad C_{A|F}=C_\psi.
\]
and
\begin{equation*}
    \mathbf g_{A|F}=
    \begin{pmatrix}
        -4g_1/3\\
        4(g_P-\chi_P)p_F
\end{pmatrix}.
\end{equation*}
The effective energy is
\begin{equation*}
    \Delta f_{A|F}^{\mathrm{eff}}(\rho;T)=\frac12r_{A|F}^{F}\rho^2+\frac14u_{A|F}^{F}\rho^4+\frac16v_{A|F}^{F}\rho^6+O(\rho^8),
\end{equation*}
where
\begin{align}
    r_{A|F}^{F}=A_\psi(T)-\frac{4g_1}{3}S_F+2(g_P-\chi_P)p_F^2,\quad u_{A|F}^{F}=B_\psi+\frac{8g_2}{9}S_F^2-\frac12\mathbf g_{A|F}^{T}H_F^{-1}\mathbf g_{A|F}.\label{eq:full-A-given-F}
\end{align}

For the smectic-first pathway \(\SmA\to\SmA_F\), with ordering coordinate \(p\), the coefficients before parent-branch relaxation are
\[
    A_{F|A}(S,\rho;T)=A_P(T)-\frac{4\lambda}{3}S+2(g_P-\chi_P)\rho^2,\quad B_{F|A}=B_P,\qquad C_{F|A}=C_P,
\]
with
\[
    \mathbf g_{F|A}=\nabla_{\mathbf y_A}A_{F|A}(\mathbf y_A;T)=
\begin{pmatrix}
-4\lambda/3\\
4(g_P-\chi_P)\rho_A
\end{pmatrix}.
\]
The effective energy is
\begin{equation*}
    \Delta f_{F|A}^{\mathrm{eff}}(p;T)=\frac12r_{F|A}^{A}p^2+\frac14u_{F|A}^{A}p^4+\frac16v_{F|A}^{A}p^6+O(p^8),
\end{equation*}
where
\begin{align}
    r_{F|A}^{A}=A_P(T)-\frac{4\lambda}{3}S_A+2(g_P-\chi_P)\rho_A^2,\quad u_{F|A}^{A}=B_P-\frac12\mathbf g_{F|A}^{T}H_A^{-1}\mathbf g_{F|A}.\label{eq:full-F-given-A}
\end{align}
The sixth-order coefficients \(v_{A|F}^{F}\) and \(v_{F|A}^{A}\) follow from \eqref{eq:246-v}, evaluated on the corresponding relaxed parent branches. The combination \(g_P-\chi_P\) controls the direct polar--smectic interaction in the fixed-wave-number reduction, while the Hessian terms describe the additional renormalization caused by parent-phase relaxation.

\paragraph{\texorpdfstring{\boldmath \(\SmA_F\to\SmC_F\)}{SmAF to SmCF}}
\label{subsec:SmAF-to-SmCF}
\hfill\break

The remaining instability of the \(\SmA_F\) branch is the tilt mode. Define
\[
    f_{\SmAF}(S,p,\rho;T):=f(S,p,\rho,0;T),\qquad\mathbf y_{AF}=(S_{AF},p_{AF},\rho_{AF}),
\]
with
\[
    \nabla_{\mathbf y}f_{\SmAF}(\mathbf y_{AF};T)=0,\qquad H_{AF}:=D_{\mathbf y}^2f_{\SmAF}(\mathbf y_{AF};T)\succ0.
\]
Using a signed tilt coordinate \(c\), with \(c^2=\sin^2\theta\), the local angular energy takes the standard even form
\[
    f(\mathbf y,c;T)=f_{\SmAF}(\mathbf y;T)+\frac12A_{C|AF}(\mathbf y)c^2+\frac14B_{C|AF}(\mathbf y)c^4,
\]
where
\[
    A_{C|AF}(\mathbf y)=2\rho^2(g_1S-g_Pp^2)-\frac{4g_2}{3}S^2\rho^4,\quad B_{C|AF}(\mathbf y)=2g_2S^2\rho^4. 
\]
There is no bare sixth-order tilt term in \eqref{eq:reduced-energy}; the effective sixth-order coefficient is generated by relaxation of the parent variables. Let
\begin{equation*}
    \mathbf g_{C|AF}:=\nabla_{\mathbf y}A_{C|AF}(\mathbf y_{AF}),\qquad\mathbf y_{C,2}^{AF}=-\frac12H_{AF}^{-1}\mathbf g_{C|AF}.
\end{equation*}
After parent-branch relaxation,
\[
    \Delta f_{C|AF}^{\mathrm{eff}}(c;T)=\frac12r_C^{AF}c^2+\frac14u_C^{AF}c^4+\frac16v_C^{AF}c^6+O(c^8),
\]
where
\begin{align*}
    &r_C^{AF}=A_{C|AF}(\mathbf y_{AF}),\qquad u_C^{AF}=B_{C|AF}(\mathbf y_{AF})-\frac12\mathbf g_{C|AF}^{T}H_{AF}^{-1}\mathbf g_{C|AF},\\
    &v_C^{AF}=\mathcal T_{AF}[\mathbf y_{C,2}^{AF},\mathbf y_{C,2}^{AF},\mathbf y_{C,2}^{AF}]+\frac32(\mathbf y_{C,2}^{AF})^TD_{\mathbf y}^2A_{C|AF}(\mathbf y_{AF})\mathbf y_{C,2}^{AF}+\frac32\nabla_{\mathbf y}B_{C|AF}(\mathbf y_{AF})\cdot\mathbf y_{C,2}^{AF},
\end{align*}
with $\mathcal T_{AF}=D_{\mathbf y}^3f_{\SmAF}(\mathbf y_{AF};T)$.
Although \(\chi_P\) does not enter the bare angular coefficients explicitly, it affects the transition through \(\mathbf y_{AF}\) and \(H_{AF}\). The transition is classified using Section~\ref{subsec:246}, provided \(v_C^{AF}>0\); a finite-amplitude solution must additionally satisfy $0<c_{\mathrm{coex}}^2\leq1$.

\subsubsection{\textbf{Competing transitions:} \texorpdfstring{\boldmath \(N\to N_F\), \(N\to\SmA\), \textbf{and} \(N\to\SmA_F\)}{N to NF, N to SmA, and N to SmAF}}
\label{sec:Near-nematic}

The simultaneous relevance of \(p\) and \(\rho\) near the nematic branch is described by the two-mode reduction of Section~\ref{subsec:two-mode}. Using the one-mode coefficients in Eqs.~\eqref{eq:near-N-polar-coefficients} and \eqref{eq:near-N-smectic-coefficients}, the quartic effective energy is
\begin{equation*}
    \Delta f_{4,N}=\frac12r_Fp^2+\frac12r_A\rho^2+\frac14u_Fp^4+\frac12wp^2\rho^2+\frac14u_A\rho^4,
\end{equation*}
where
\begin{equation}
    w(T)=2(g_P-\chi_P)-\frac{8\lambda g_1}{9H_N(T)}. \label{eq:w}
\end{equation}
Accordingly, the quartic determinant \(D\) defined in \eqref{eq:two-mode-D} becomes
\begin{equation}
    D_N(T):=u_F(T)u_A(T)-w^2(T). \label{eq:near-N-D}
\end{equation}
For \(D_N\neq0\), the mixed branch is
\begin{equation}
    p^2=\frac{wr_A-u_A r_F}{D_N},\qquad\rho^2=\frac{wr_F-u_F r_A}{D_N}.\label{eq:near-N-mixed-branch}
\end{equation}
It is locally stable at quartic order when $u_F>0$, $u_A>0$, $D_N>0$, $p^2>0$ and $\rho^2>0$. The transverse coefficients for sequential ordering close to the nematic branch are
\begin{align}
    r_{A|F}&=r_A-\frac{w}{u_F}r_F,&u_{A|F}&=u_A-\frac{w^2}{u_F}=\frac{D_N}{u_F},\label{eq:near-N-transverse-A-given-F}\\
    r_{F|A}&=r_F-\frac{w}{u_A}r_A,&u_{F|A}&=u_F-\frac{w^2}{u_A}=\frac{D_N}{u_A}.\label{eq:near-N-transverse-F-given-A}
\end{align}
These expressions apply only while both amplitudes remain small. Once the first daughter phase is developed, they are replaced by the fully relaxed coefficients in Eqs.~\eqref{eq:full-A-given-F} and \eqref{eq:full-F-given-A}.

\begin{figure}[H]
    \centering
    \begin{subfigure}[t]{0.325\linewidth}
        \centering
        \phantomsubcaption
        \label{fig:competition-cooperative}
        \begin{overpic}[width=\linewidth]{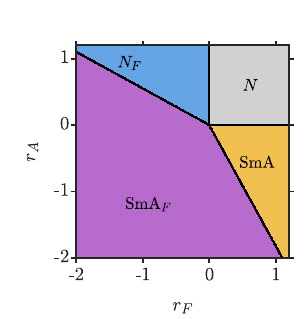}
            \put(3,93){(a)}
        \end{overpic}
    \end{subfigure}
    \hfill
    \begin{subfigure}[t]{0.325\linewidth}
        \centering
        \phantomsubcaption
        \label{fig:competition-weak}
        \begin{overpic}[width=\linewidth]{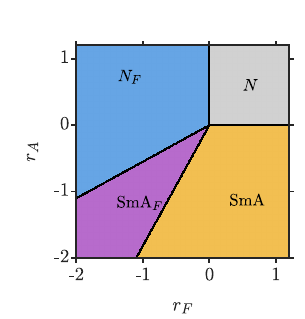}
            \put(3,93){(b)}
        \end{overpic}
    \end{subfigure}
    \hfill
    \begin{subfigure}[t]{0.325\linewidth}
        \centering
        \phantomsubcaption
        \label{fig:competition-strong}
        \begin{overpic}[width=\linewidth]{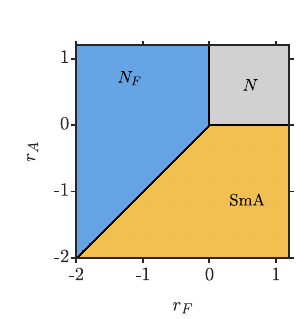}
            \put(3,93){(c)}
        \end{overpic}
    \end{subfigure}
    \vspace{-0.8em}
    \caption{Local phase maps of the quartic polar--smectic model in the $(r_F,r_A)$ plane, with $u_F=u_A=1$: (a) cooperative coupling, $D_N>0$ and $w<0$; (b) weak competition, $D_N>0$ and $w>0$; (c) strong competition, $D_N<0$ and $w>0$.}
    \label{fig:competition}
\end{figure}

In the quartic regime \(u_F>0\) and \(u_A>0\), the sign of the determinant \(D_N\) in \eqref{eq:near-N-D}, together with the sign of \(w\) in \eqref{eq:w}, determines the local competition regime in the \((r_F,r_A)\) plane. Fig.~\ref{fig:competition} illustrates three representative cases. These are coefficient-space maps rather than temperature phase diagrams; a cooling trajectory corresponds to a trajectory through the \((r_F,r_A)\) plane.

In Fig.~\ref{fig:competition-cooperative}, \(D_N>0\) and \(w<0\), corresponding to cooperative coupling. The nematic phase \(N\) is locally stable for \(r_F>0\) and \(r_A>0\), since both ordering modes have positive quadratic curvature. If \(r_F<0\) while \(r_{A|F}>0\), the polar mode orders first and the stable branch is \(N_F\); similarly, \(r_A<0\) together with \(r_{F|A}>0\) gives the \(\SmA\) branch. When both transverse coefficients in Eqs.~\eqref{eq:near-N-transverse-A-given-F} and \eqref{eq:near-N-transverse-F-given-A} become negative, namely $r_{A|F}<0$ and $r_{F|A}<0$, the locally stable mixed branch is \(\SmA_F\). Because \(w<0\), the presence of either order lowers the transverse quadratic coefficient of the other, so the two ordering tendencies promote one another.

Fig.~\ref{fig:competition-weak} shows the weakly competitive regime, \(D_N>0\) and \(w>0\). The mixed \(\SmA_F\) phase therefore occupies the region \(r_{A|F}<0\) and \(r_{F|A}<0\), bounded by $r_A=w r_F/u_F$ and $r_F=wr_A/u_A$. For \(u_F=u_A=1\), increasing \(w\) within the range \(0<w<1\) brings these two boundaries closer together and hence reduces the \(\SmA_F\) region.

In Fig.~\ref{fig:competition-strong}, \(D_N<0\) and \(w>0\), corresponding to strong competition. Since the quartic form is then indefinite, an interior mixed stationary branch cannot be a local minimum, and no locally stable \(\SmA_F\) region occurs at quartic order. Where the \(N_F\) and \(\SmA\) branches coexist as locally stable states, the equilibrium boundary is obtained by comparing their reduced free energies, $\Delta F_{N_F}=-r_F^2/(4u_F)$ and $\Delta F_{\SmA}=-r_A^2/(4u_A)$. Their equality gives $r_F^2/u_F=r_A^2/u_A$, indicating $r_A=\sqrt{u_A/u_F}r_F$ when $r_F<0$ and $r_A<0$. For \(u_F=u_A=1\), this reduces to \(r_A=r_F\), which is the boundary between the \(N_F\) and \(\SmA\) regions in Fig.~\ref{fig:competition-strong}.

The strongly cooperative regime \(D_N<0\), \(w<0\) is not shown in Fig.~\ref{fig:competition}; there the quartic form is unstable along a mixed direction, and a possible finite-amplitude \(N\to\SmA_F\) transition must be treated by the directional \(2\!-\!4\!-\!6\) reduction of Section~\ref{subsec:two-mode}.

\subsubsection{\textbf{Global selection of polar--smectic ordering pathways}}
\label{subsec:global-selection-amplitude}
Here the two ordering coordinates are the polar and smectic amplitudes, $(\xi_1,\xi_2)=(p,\rho)$ giving the $N$, $N_F$, $\SmA$, and $\SmA_F$ branches. Candidate transitions are classified by the coefficients derived above, while the realized pathway is selected by comparing the fully relaxed free energies of the locally stable and physically admissible amplitude
branches:
\begin{equation}\label{eq:amplitude-global-selection}
    \beta_{\rm eq}(T)\in\operatorname*{arg\,min}_{\substack{\beta\in\{N,N_F,\SmA,\SmA_F\}\\ \beta\ {\rm locally\ stable\ and\ admissible}}}F_\beta(T).
\end{equation}

\paragraph{\texorpdfstring{\boldmath\(N_F\to\SmA\)}{NF to SmA}}
\hfill\break

When \(w>0\) and \(D_N<0\), the mixed quartic branch is unstable and the \(N_F\) and \(\SmA\) branches may coexist as locally stable minima. At quartic order, their switching temperature \(T_{F\to A}^{(4)}\) is estimated from
\begin{equation}
    \frac{r_F^2(T_{F\to A}^{(4)})}{u_F(T_{F\to A}^{(4)})}=\frac{r_A^2(T_{F\to A}^{(4)})}{u_A(T_{F\to A}^{(4)})},\label{eq:NF-SmA-coexistence}
\end{equation}
subject to the bistability conditions $r_F<0$, $r_A<0$ with $r_{A|F}>0$, $r_{F|A}>0.$ The coexistence temperature of the fully relaxed \(N_F\) and \(\SmA\) branches is determined by: $F_{N_F}(T_{F\to A})=F_{\SmA}(T_{F\to A})$, $H_F(T_{F\to A})\succ0$ and $H_A(T_{F\to A})\succ0$. Whether this coexistence corresponds to the equilibrium transition is determined by the global free-energy comparison \eqref{eq:amplitude-global-selection}. When the two coexisting minima are distinct, the transition is first order because \(p\) and \(\rho\) change discontinuously across the branch switch.

\subsubsection{Assembled ferroelectric cooling sequences}
\label{subsec:assembled-ferro-smectic-sequences}

Combining the amplitude pathways of Section~\ref{subsec:global-selection-amplitude} with the tilt reduction of Section~\ref{subsec:SmAF-to-SmCF} gives the principal cooling sequences $N\to N_F\to\SmAF\to\SmCF$, $N\to\SmA\to\SmAF\to\SmCF$ and $N\to\SmAF\to\SmCF$, corresponding, respectively, to polar-first, smectic-first, and direct mixed amplitude ordering. If the tilt instability is not reached, the corresponding sequence terminates at \(\SmAF\).

Strong polar--smectic competition can additionally produce $N\to N_F\to\SmA\to\SmAF\to\SmCF$,  where the first-order \(N_F\to\SmA\) branch switch is followed by the \(\SmA\to\SmAF\) onset. This pathway illustrates how competition can alter the intermediate phase sequence without changing the final \(\SmCF\) phase.

These sequences are representative rather than exhaustive. Other pathways, such as a direct \(\SmA\to\SmCF\) transition involving simultaneous polarization and tilt ordering, can be constructed within the same modular framework.

\subsection{Numerical results and experimental comparison}
\label{sec:numerical-calibration}

The analytical reductions developed above determine the local transition type and the admissible competing branches. We now illustrate how these local criteria combine with the global branch-selection rule to generate equilibrium phase sequences. Representative phase diagrams and cooling trajectories are presented first, followed by a fitting to the experimental results of ferroelectric liquid crystal reported by Strachan \textit{et al.}~\cite{Strachan2025}. Numerical implementation and fitting details are collected in Appendix~\ref{app:numerical-calibration}.

\subsubsection{\textbf{Phase diagrams}}
\label{subsec:numerical-phase-diagrams}

\paragraph{Global numerical branch selection.}
For each temperature and control parameter, all locally stable and physically admissible candidate branches within the reduced homogeneous ansatz are fully relaxed, and the equilibrium branch \(\beta_{\rm eq}(T)\) is selected by comparing their fully relaxed free energies. Thus, the numerical calculation implements the modular phase-selection procedure of Section~\ref{subsec:branch-adaptive-reduction} within the adopted reduced model. The numerical branch-selection procedure and parameter sets are summarized in Appendix~\ref{subsec:numerical-branch-selection}.

\par\medskip
\noindent\textbf{Polar--smectic ordering phase diagrams}
\label{subsubsec:polar-smectic-phase-diagrams}
\par\smallskip

We first examine how the relative ordering tendency and the effective polar--smectic interaction affect the equilibrium pathways. The offset $\Delta T=T_P^*-T_\psi^*$ controls the relative polar and smectic instabilities, whereas \(g_P\) and \(\chi_P\) enter the effective mixed coupling \(w(T)\) in \eqref{eq:w} and thereby modify the competition between the two modes. The parameter sets used in the scans are listed in Table~\ref{tab:generic-phase-diagram-parameters}.

\begin{figure}[htbp]
    \centering
    \begin{subfigure}[t]{0.485\linewidth}
        \centering
        \caption{Sequential polar-first and smectic-first pathways.}
        \includegraphics[width=\linewidth]{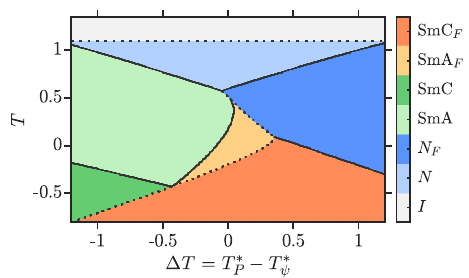}
        
        \label{fig:scan-deltaT}
    \end{subfigure}
    \hfill
    \begin{subfigure}[t]{0.49\linewidth}
        \centering
        \caption{Direct formation of the mixed \(\SmAF\) phase.}
        \includegraphics[width=\linewidth]{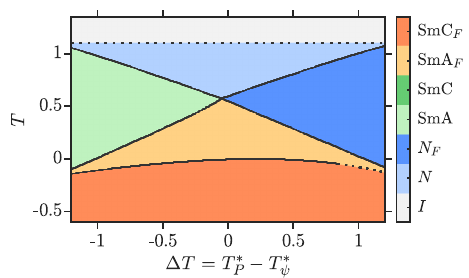}
        
        \label{fig:scan-deltaT2}
    \end{subfigure}
    \caption{Equilibrium phase diagrams in the \((\Delta T,T)\) plane for two representative polar--smectic coupling regimes. Colors identify the globally stable phases. Black solid and dotted lines denote continuous and first-order phase boundaries, respectively.}
    \label{fig:global-amplitude-phase-selection}
\end{figure}
Fig.~\ref{fig:global-amplitude-phase-selection} shows how the equilibrium cooling pathway changes with the relative polar and smectic ordering temperature, \(\Delta T=T_P^*-T_\psi^*\). In Fig.~\ref{fig:scan-deltaT}, for the relatively competitive choice \(g_P=0.60\) and \(\chi_P=0.08\), varying \(\Delta T\) from negative to positive changes the sequence from \(N\to\SmA\to\SmA_F\) to \(N\to N_F\to\SmA_F\), with a competition region between the two pathways. For the more cooperative choice \(g_P=0.35\) and \(\chi_P=0.25\) in Fig.~\ref{fig:scan-deltaT2}, a direct \(N\to\SmA_F\) region also appears, where polar and smectic order develop simultaneously without an intervening \(N_F\) or \(\SmA\) phase.

\begin{figure}[htbp]
    \centering
    \begin{subfigure}[t]{0.485\linewidth}
        \centering
        \caption{$I\to N\to N_F\to\mathrm{SmA}_F\to\mathrm{SmC}_F$.}
        \includegraphics[width=\linewidth]{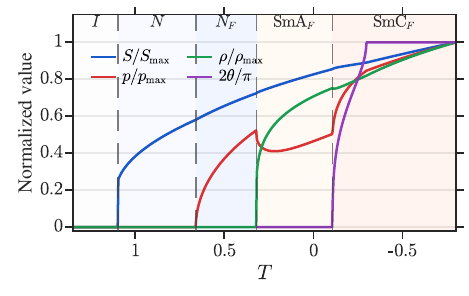}  
        \label{fig:sequence-a}
    \end{subfigure}
    \hfill
    \begin{subfigure}[t]{0.485\linewidth}
        \centering
        \caption{$I\to N\to\mathrm{SmA}\to \mathrm{SmA}_F\to\mathrm{SmC}_F$.}
        \includegraphics[width=\linewidth]{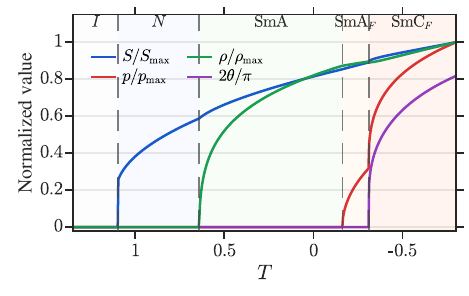}
        \label{fig:sequence-b}
    \end{subfigure}
    \vspace{-1.5em}
    \begin{subfigure}[b]{0.485\linewidth}
        \centering
        \caption{$I\to N\to N_F\to\mathrm{SmA}\to\mathrm{SmA}_F\to\mathrm{SmC}_F$.}
        \includegraphics[width=\linewidth]{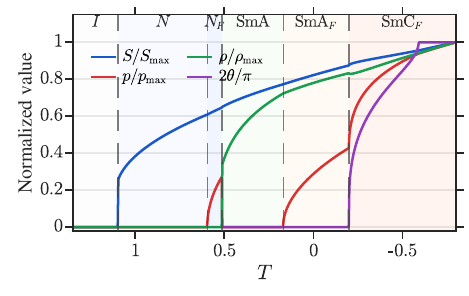}
        \label{fig:sequence-c}
    \end{subfigure}
    \hfill
    \begin{subfigure}[b]{0.485\linewidth}
        \centering
        \caption{$I\to N\to\mathrm{SmA}_F\to\mathrm{SmC}_F$.}
        \includegraphics[width=\linewidth]{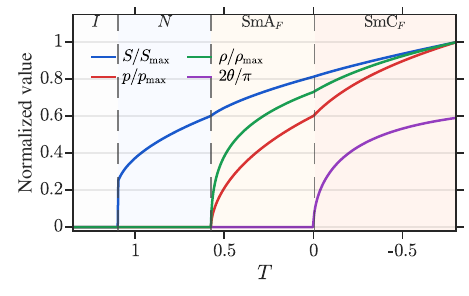}
        \label{fig:sequence-d}
    \end{subfigure}    
    \vspace{-1.0em}
    \caption{Normalized order-parameter trajectories for four cooling sequences. Shaded regions indicate the equilibrium phases, and dashed vertical lines mark the phase boundaries.}
    \label{fig:four-cooling-sequences}
\end{figure}

Fig.~\ref{fig:four-cooling-sequences} shows four representative cooling trajectories selected from the phase diagrams in Fig.~\ref{fig:global-amplitude-phase-selection}. Figs.~\ref{fig:sequence-a}--\ref{fig:sequence-c} correspond to three cuts through Fig.~\ref{fig:scan-deltaT}, representing respectively the polar-first, smectic-first, and competing-branch regimes.

Fig~\ref{fig:sequence-a} shows a representative polar-first cooling trajectory for \(\Delta T=0.15\) in Fig~\ref{fig:scan-deltaT}. As \(T\) decreases from about \(1.35\) to \(-0.8\), the system first undergoes a first-order \(I\to N\) transition at \(T\simeq1.10\), where the normalized nematic order \(S/S_{\max}\) jumps from zero to about \(0.26\). At \(T\simeq0.66\), the continuous \(N\to N_F\) transition occurs. On further cooling to \(T\simeq0.32\), a first-order \(N_F\to\SmA_F\) transition occurs, where the normalized smectic amplitude \(\rho/\rho_{\max}\) jumps from zero to about \(0.35\), accompanied by a discontinuous decrease in \(p/p_{\max}\). Finally, at \(T\simeq-0.10\), the first-order \(\SmA_F\to\SmC_F\) transition occurs, with the normalized tilt \(2\theta/\pi\) jumping from zero to about \(0.30\). The resulting polar-first sequence \(I\to N\to N_F\to\SmA_F\to\SmC_F\) follows the same ordering hierarchy reported experimentally by Strachan \textit{et al.}~\cite{Strachan2025}.

Fig.~\ref{fig:sequence-b} shows a smectic-first cooling trajectory for \(\Delta T=-0.20\) in Fig.~\ref{fig:scan-deltaT}. The initial \(I\to N\) transition is first order, as in Fig.~\ref{fig:sequence-a}. At \(T\simeq0.64\), a continuous \(N\to\SmA\) transition occurs, with the smectic amplitude \(\rho/\rho_{\max}\) emerging continuously from zero. Polar order appears only upon further cooling, through a first-order \(\SmA\to\SmA_F\) transition at \(T\simeq-0.17\), where \(p/p_{\max}\) acquires a finite jump from zero. Upon further cooling, the \(\SmA_F\to\SmC_F\) transition introduces the tilt, as also seen in Fig.~\ref{fig:sequence-a}. The resulting sequence is \(I\to N\to\SmA\to\SmA_F\to\SmC_F\), with smectic order developing before polar order.

Fig.~\ref{fig:sequence-c} shows a strong-competition pathway for \(\Delta T=0.00\) in Fig.~\ref{fig:scan-deltaT}. The initial first-order \(I\to N\) and continuous \(N\to N_F\) transitions are similar to those in Fig.~\ref{fig:sequence-a}, so polar order again develops first. The subsequent behavior is different: at \(T\simeq0.52\), a first-order \(N_F\to\SmA\) transition occurs, with \(p/p_{\max}\) dropping discontinuously from about \(0.25\) to zero while \(\rho/\rho_{\max}\) jumps from zero to about \(0.30\). This transition is a switch between competing locally stable \(N_F\) and \(\SmA\) branches, determined by their fully relaxed free energies. Upon further cooling, a first-order \(\SmA\to\SmA_F\) transition at \(T\simeq0.17\) restores polar order with a finite jump in \(p\), followed by the \(\SmA_F\to\SmC_F\) transition at \(T\simeq-0.20\). The resulting sequence is \(I\to N\to N_F\to\SmA\to\SmA_F\to\SmC_F\).

Finally, Fig.~\ref{fig:sequence-d} shows the cooling trajectory for direct mixed ordering for \(\Delta T=-0.05\) in Fig~\ref{fig:scan-deltaT2}. The initial first-order \(I\to N\) transition is similar to those in the preceding panels. In contrast to Figs.~\ref{fig:sequence-a}--\ref{fig:sequence-c}, the nematic phase then transforms directly into \(\SmA_F\) at \(T\simeq0.57\): the polar and smectic amplitudes \(p\) and \(\rho\) both emerge continuously from zero. Upon further cooling, the continuous \(\SmA_F\to\SmC_F\) transition occurs near \(T\simeq0\), where the tilt develops continuously from zero. The resulting sequence is \(I\to N\to\SmA_F\to\SmC_F\), without an intermediate \(N_F\) or \(\SmA\) phase. Such direct development of ferroelectric smectic order from a nematic phase has also been reported experimentally \cite{Weglowska2025}.

\begin{figure}[htbp]
    \centering
    \includegraphics[width=0.5\linewidth]{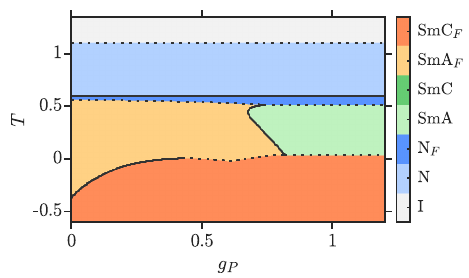}
    \vspace{-0.8em}
    \caption{Equilibrium phase diagram in the \((g_P,T)\) plane.}
    \label{fig:scan-gP}
\end{figure}

Fig~\ref{fig:scan-gP} isolates the effect of the polar--smectic coupling \(g_P\) by varying \(g_P\) at fixed \(\chi_P=0.25\) and \(\Delta T=0\). As \(g_P\) increases, the stability region of the mixed \(\SmA_F\) phase is progressively suppressed, while the \(N_F\) and \(\SmA\) single-mode regions become more prominent. This trend is consistent with Eq.~\eqref{eq:w}, since \(\partial_{g_P}w=2>0\): increasing \(g_P\) raises the effective mixed quartic coupling and therefore enhances competition between polar and smectic order. Conversely, Eq.~\eqref{eq:w} gives \(\partial_{\chi_P}w=-2\), so increasing \(\chi_P\) favors the opposite, more cooperative tendency, although \(\chi_P\) is not independently scanned in Fig.~\ref{fig:scan-gP}. This is also consistent with the contrast between Figs.~\ref{fig:scan-deltaT} and \ref{fig:scan-deltaT2}, where the more cooperative parameter set has smaller \(g_P\) and larger \(\chi_P\).

\subsubsection{Experimental fitting}\label{subsec:experiment-cal}

Strachan \textit{et al.}~\cite{Strachan2025} reported the cooling
sequence $I\to N\to N_F\to\SmA_F\to\SmC_F,$ with transition temperatures \(263^\circ\mathrm C\), \(185^\circ\mathrm C\), \(167^\circ\mathrm C\), and \(115^\circ\mathrm C\), respectively. The \(N\to N_F\) transition is continuous, the \(N_F\to\SmA_F\) transition is weakly first order, and the \(\SmA_F\to\SmC_F\) transition is continuous. In addition to the transition temperatures and phase assignments, optical birefringence was measured across the liquid-crystalline range, while X-ray diffraction provided the smectic layer spacing in the \(\SmA_F\) and \(\SmC_F\) phases. The birefringence shows a dip near the \(N\to N_F\) transition and a small step at \(N_F\to\SmA_F\), whereas the layer spacing is nearly constant in \(\SmA_F\) and decreases upon cooling through \(\SmC_F\).

\begin{figure}[H]
    \centering
    \includegraphics[width=0.85\linewidth]{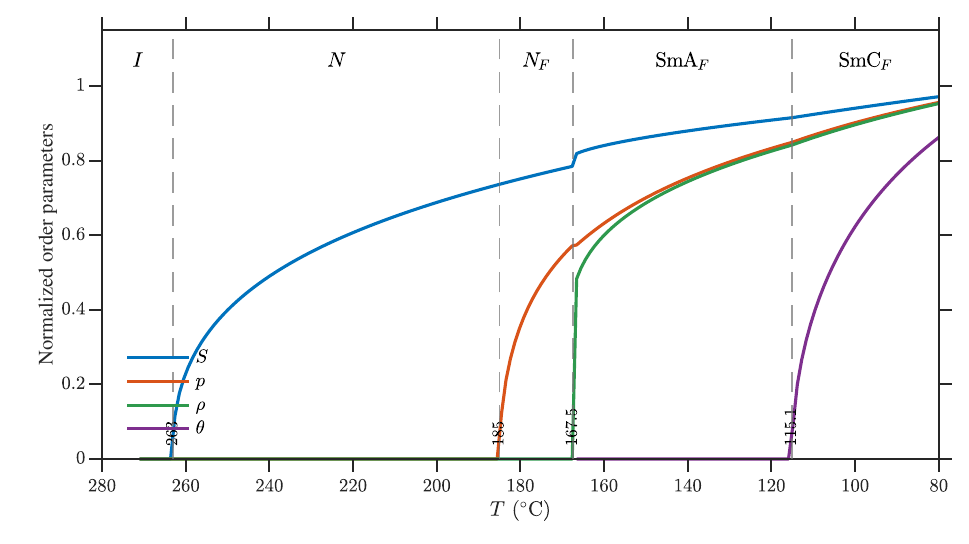}
    \vspace{-1.5em}
    \caption{Normalized equilibrium order parameters as functions of temperature for the fitted model.}
    \label{fig:order-emerge}
\end{figure}

We fit the reduced model to these experimental constraints using the least-squares procedure described in Appendix~\ref{app:numerical-calibration}. The resulting equilibrium order parameters are shown in Fig.~\ref{fig:order-emerge}.

Fig~\ref{fig:order-emerge} gives the fitted cooling sequence \(I\to N\to N_F\to\SmA_F\to\SmC_F\). At \(263.00^\circ\mathrm C\), the \(I\to N\) transition is weakly first order and \(S\) acquires a finite jump. At \(185.00^\circ\mathrm C\), the continuous \(N\to N_F\) transition occurs, with \(p\) emerging continuously from zero. On further cooling to \(167.48^\circ\mathrm C\), the first-order \(N_F\to\SmA_F\) transition is marked by a finite onset of \(\rho\). Across this transition, \(S\) increases while \(p\) shows a small decrease. These changes reflect the coupling among the already developed orders. The onset of smectic order is favored by the \(-g_1S\rho^2\)-type nematic--smectic coupling, which tends to enhance \(S\), whereas the positive mixed term \((g_P-\chi_P)p^2\rho^2\) weakly suppresses \(p\) as \(\rho\) develops. Finally, at \(115.12^\circ\mathrm C\), the \(\SmA_F\to\SmC_F\) transition is continuous, with the tilt \(\theta\) emerging continuously from zero while \(S\), \(p\), and \(\rho\) remain finite.

\begin{table}[H]
\centering
\small
\setlength{\tabcolsep}{3pt}
\renewcommand{\arraystretch}{1.20}
\caption{Comparison of the experimental and fitted transition temperatures and transition types. Temperatures and errors are in \(^{\circ}\mathrm C\).}
\label{tab:transition_character}

\begin{tabular*}{\textwidth}{@{\extracolsep{\fill}}cccc
    >{\raggedright\arraybackslash}p{0.20\textwidth}
    >{\raggedright\arraybackslash}p{0.18\textwidth}@{}}
\toprule
Transition & \(T_{\mathrm{exp}}\) & \(T_{\mathrm{model}}\) & Error & Experiment & Model \\
\midrule
\(I\to N\) & \(263\) & \(263.00\) & \(0.00\) & Not specified & Weakly first-order \\
\(N\to N_F\) & \(185\) & \(185.00\) & \(0.00\) & Continuous & Continuous \\
\(N_F\to\SmA_F\) & \(167\) & \(167.48\) & \(+0.48\) & Weakly first-order & First-order \\
\(\SmA_F\to\SmC_F\) & \(115\) & \(115.12\) & \(+0.12\) & Continuous & Continuous \\
\bottomrule
\end{tabular*}
\end{table}
Table~\ref{tab:transition_character} compares the fitted transition temperatures and transition types with experiment. The \(I\to N\) and \(N\to N_F\) transition temperatures are reproduced exactly because \(263^\circ\mathrm C\) and \(185^\circ\mathrm C\) are imposed as transition conditions in the fitting and are used to determine the corresponding derived parameters, see Appendix~\ref{subsec:calibration-model}. By contrast, the \(N_F\to\SmA_F\) and \(\SmA_F\to\SmC_F\) transition temperatures are determined from the fully relaxed branch crossing and the tilt instability, respectively. Their fitted values, \(167.48^\circ\mathrm C\) and \(115.12^\circ\mathrm C\), differ from the experimental values by only \(0.48^\circ\mathrm C\) and \(0.12^\circ\mathrm C\). The fitted transition types are also consistent with experiment: \(N\to N_F\) and \(\SmA_F\to\SmC_F\) are continuous, whereas \(N_F\to\SmA_F\) is first order, corresponding to the weakly first-order transition reported experimentally.

\begin{figure}[htbp]
    \centering
    \begin{subfigure}[t]{0.485\linewidth}
        \centering
        \caption{Birefringence.}
        \includegraphics[width=\linewidth]{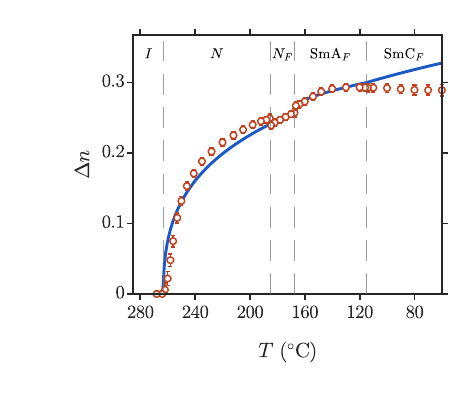}
        \label{fig:delta-n}
    \end{subfigure}
    \hfill
    \begin{subfigure}[t]{0.485\linewidth}
        \centering
        \caption{Smectic layer spacing.}
        \includegraphics[width=\linewidth]{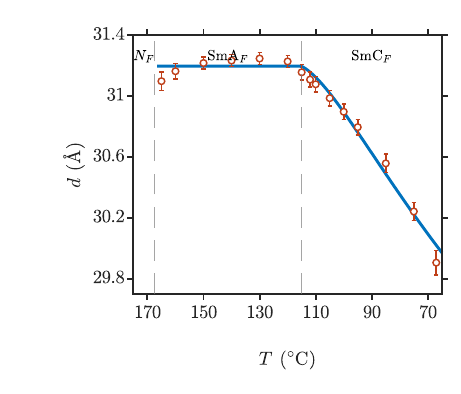}
        \label{fig:d-a}
    \end{subfigure}
    \vspace{-2.0em}
    \caption{Comparison between the fitted model and experimental observables: (a) optical birefringence and (b) smectic layer spacing. Red dots denote the experimental data.}
    \label{fig:cal}
\end{figure}

We next compare the fitted equilibrium solution with the experimental observables in Fig.~\ref{fig:cal}, where the reconstructed birefringence and smectic layer spacing are shown together with the measured data. Fig~\ref{fig:delta-n} shows the birefringence, which increases overall upon cooling and reflects the growth of nematic orientational order. The remaining quantitative deviations, including the shallow anomaly near the \(N\to N_F\) transition and the low-temperature plateau, are not captured by the simplified relation \(\Delta n=n_{\mathrm{scale}}S\). A possible reason for the low-temperature plateau is that the optical anisotropy is no longer determined by the nematic order alone: polar, smectic, and tilt ordering may provide additional contributions that partly compensate the further growth of \(S\). Such effects are not included in the present linear mapping. Fig~\ref{fig:d-a} shows the smectic layer spacing, reconstructed from the equilibrium tilt \(\theta(T)\) using the relation described in Appendix~\ref{subsec:calibration-model}. The layer spacing remains approximately constant in the \(\SmA_F\) phase and decreases continuously after the onset of tilt in the \(\SmC_F\) phase, in agreement with the main structural trend observed experimentally.

\section{Conclusions and outlook}
\label{sec:conclusion}

We have developed a modular Landau framework for coupled phase transitions from a locally stable parent phase. A central feature of the construction is that the noncritical parent variables are relaxed before the emerging instability is classified, rather than being frozen at their parent-phase values. This relaxation systematically renormalizes the effective Landau coefficients and can have direct consequences for the transition behavior. In the asymmetric \(2\!-\!3\!-\!4\) reduction, it lowers the effective quartic coefficient, shifting the coexistence and spinodal conditions and increasing the order-parameter jump relative to the frozen-parent description under the usual temperature-crossing assumptions. In the symmetry-preserving \(2\!-\!4\!-\!6\) reduction, relaxation can drive the effective quartic coefficient through the tricritical value and thereby convert a transition that is continuous in the frozen-parent description into a first-order one. For two coupled ordering modes, the same mechanism renormalizes the quartic coupling matrix and hence enters the competition between single-mode and mixed ordering.

These one- and two-mode reductions form reusable transition modules describing asymmetric, symmetry-preserving, mixed, and competing transitions. After a daughter phase is selected, it becomes the parent branch for the next transition. Repeating this branch update assembles the local modules into a complete equilibrium phase sequence, while comparison of fully relaxed branch free energies selects the realized pathway. The construction depends on the local symmetry, stability, and transition topology rather than on a particular material system or choice of order parameters.

Ferroelectric liquid crystals provide a concrete application involving coupled orientational, polar, smectic, and tilt ordering. Using a symmetry-consistent \(Q+P+\psi\) free energy within the adopted constrained homogeneous ansatz, the general reductions can be applied successively to the individual liquid-crystal transitions and assembled into several competing cooling pathways. The numerical results show how the relative ordering temperatures and polar--smectic couplings control the selection among these pathways. A single fitted thermodynamic parameter set reproduces the experimental phase sequence and transition types reported by Strachan \textit{et al.}~\cite{Strachan2025}, together with the main trends of the orientational and polar order. The smectic layer spacing is described separately by a low-order post-fit reconstruction based on the calculated equilibrium tilt. This application therefore provides a direct test of the modular construction in a system with several interacting forms of order.


Looking forward, the modular construction can first be extended within liquid-crystal systems to more complex combinations of orientational, polar, antipolar, smectic, and tilt ordering. Examples include polar--antipolar and modulated ordering in ferroelectric and antiferroelectric liquid crystals, including the \(\mathrm{SmZ}_A\), \(N_{TBF}\), and related ferroelectric smectic phases. The direct continuous \(N\to\SmC\) transition is another particularly interesting case, because the usual scalar \((\rho,\theta)\) parametrization becomes singular as \(\rho\to0\); a nonsingular ordering coordinate combining smectic amplitude and director--layer orientation would provide a useful test of the modular reduction under a different choice of order parameters \cite{ChenLubensky1976,MukamelHornreich1980}. Beyond liquid crystals, the same parent-relaxation and coupled-mode construction can be applied to structural--polar ordering in ferroelectric materials, magnetic--polar ordering in multiferroics, and competing magnetic--superconducting orders in unconventional superconductors \cite{VanderbiltCohen2001,LiCrossChen2005,Mostovoy2006,FernandesSchmalian2010,BruceAharony1975}. These systems provide natural settings in which several symmetry-breaking modes coexist or compete and in which successive transitions cannot in general be analyzed independently. A further extension is to spatially heterogeneous phase transitions. Incorporating gradient and elastic contributions would connect the present homogeneous phase-selection framework with interfaces, domain walls, defects, modulated structures, and confinement in liquid crystals and other ferroic materials. Together with reductions involving more than two simultaneously relevant ordering modes, this would extend the modular construction from bulk phase sequences to spatially structured and higher-dimensional multicritical phenomena.

\appendix
\section{Numerical implementation and fitting details}
\label{app:numerical-calibration}
\subsection{Numerical equilibrium selection}
\label{subsec:numerical-branch-selection}

For each temperature \(T\) and scanned control parameter \(\mu\), we minimize the reduced homogeneous free energy within each candidate phase. Denoting the corresponding locally stable equilibrium branch by \(\beta\), its relaxed free energy is written as $F_\beta(T,\mu)$. The thermodynamic equilibrium phase is then determined by direct comparison of the relaxed branch free energies,
\begin{equation}
    \beta_{\mathrm{eq}}(T;\mu)\in\operatorname*{arg\,min}_{\beta\in\mathcal B_{\mathrm{loc}}(T;\mu)}F_\beta(T;\mu),\label{eq:numerical-global-selection}
\end{equation}
where \(\mathcal B_{\mathrm{loc}}(T;\mu)\) denotes the set of locally stable and physically admissible branches. Higher-energy local minima, when present, are interpreted as metastable states.

A transition is classified as continuous when the equilibrium daughter branch emerges from the parent branch with a vanishing ordering amplitude. A first-order transition occurs when two distinct locally stable branches exchange global stability with a finite discontinuity in the relevant order parameter.

The generic phase diagrams in Figs.~\ref{fig:scan-deltaT}, \ref{fig:scan-deltaT2}, and \ref{fig:scan-gP} are obtained by applying \eqref{eq:numerical-global-selection} throughout the corresponding \((T,\mu)\)-parameter plane. Black solid and dotted lines denote continuous and first-order transitions, respectively.

The parameter sets used for the numerical phase-diagram scans are listed in Table~\ref{tab:generic-phase-diagram-parameters}.

\begin{table}[H]
\centering
\small
\renewcommand{\arraystretch}{1.18}
\setlength{\tabcolsep}{4pt}
\caption{Model parameters used for the numerical phase-diagram scans.}
\label{tab:generic-phase-diagram-parameters}
\resizebox{\textwidth}{!}{
\begin{tabular}{
    l
    cccc
    ccc
    ccc
    ccc
    ccc
}
\toprule

& \multicolumn{4}{c}{Nematic} & \multicolumn{3}{c}{Polar} & \multicolumn{3}{c}{Smectic} & \multicolumn{3}{c}{Couplings} & \multicolumn{3}{c}{Control parameters}\\

\cmidrule(lr){2-5}
\cmidrule(lr){6-8}
\cmidrule(lr){9-11}
\cmidrule(lr){12-14}
\cmidrule(lr){15-17}

Fig & \(a_Q\) & \(T_Q^*\) & \(B_Q\) & \(C_Q\) & \(a_P\) & \(B_P\) & \(C_P\) & \(a_\psi\) & \(B_\psi\) & \(C_\psi\) & \(\lambda\) & \(g_1\) & \(g_2\) & \(\Delta T^*\) & \(g_P\) & \(\chi_P\)\\

\midrule

Fig~\ref{fig:scan-deltaT} & 1.0 & 1.0 & 2.8 & 3.0 & 1.0 & 2.0 & 1.0 & 1.0 & 0.20 & 1.0 & 0.25 & 0.20 & 0.30 & \([-1.20,1.20]\) & 0.60 & 0.08\\

Fig~\ref{fig:scan-deltaT2}& 1.0 & 1.0 & 2.8 & 3.0 & 1.0 & 2.0 & 1.0 & 1.0 & 0.20 & 1.0 & 0.25 & 0.20 & 0.30 & \([-1.20,1.20]\) & 0.35 & 0.25 \\

Fig~\ref{fig:scan-gP} & 1.0 & 1.0 & 2.8 & 3.0 & 1.0 & 2.0 & 1.0 & 1.0 & 0.20 & 1.0 & 0.25 & 0.20 & 0.30 & 0 & \([0,1.20]\) & 0.25 \\

\bottomrule
\end{tabular}}
\end{table}
For the \(\Delta T^*\)-scans, \(T_P^*\) and \(T_\psi^*\) are varied symmetrically about \(0.35\), so that
\[
    T_P^*=0.35+\frac{\Delta T^*}{2},\qquad T_\psi^*=0.35-\frac{\Delta T^*}{2}.
\]
For the \(g_P\)-scan, \(\Delta T^*=0\), with \(T_P^*=T_\psi^*=0.35\).

\subsection{Experimental fitting and parameter determination}
\label{subsec:calibration-model}

We fit the reduced homogeneous model to the experimental results reported by Strachan et al.~\cite{Strachan2025}. The fitting combines imposed transition conditions, experimental observables, and auxiliary constraints used to preserve the observed transition type. The experimental cooling sequence, transition temperatures, and transition types are summarized in Table~\ref{tab:transition_character}. In addition, the birefringence is used to constrain the orientational order, while representative polarization values are used to constrain the magnitude and temperature dependence of the polar order. The X-ray layer-spacing data are not included in the thermodynamic fitting; instead, the layer spacing is reconstructed separately after the thermodynamic parameter set has been determined.

\paragraph{Experimental data used in the fitting.}
The birefringence data used in the fit are obtained from the optical measurements reported in~\cite{Strachan2025}. We denote the digitized data by $\{T_i,\Delta n_i^{\mathrm{exp}}\}_{i=1}^{N_{\Delta n}},$ where \(T_i\) is the temperature of the \(i\)-th data point and \(\Delta n_i^{\mathrm{exp}}\) is the corresponding experimental birefringence. For the staged fitting, the birefringence data are grouped according to the experimental phase intervals defined by the transition temperatures in Table~\ref{tab:transition_character}. The model prediction is related to the equilibrium nematic amplitude through $\Delta n_{\mathrm{mod}}(T)= n_{\mathrm{scale}} S_{\mathrm{eq}}(T),$ with $n_{\mathrm{scale}}=0.330$, where \(S_{\mathrm{eq}}(T)\) denotes the equilibrium orientational amplitude of the thermodynamically selected branch. The factor \(n_{\mathrm{scale}}\) is kept fixed throughout the fitting.

The available polarization data are too sparse for a pointwise fit as a function of temperature. We therefore relate the experimental spontaneous polarization to the equilibrium polar amplitude through $P_s(T)=P_{\mathrm{scale}}\,p_{\mathrm{eq}}(T)$, where \(p_{\mathrm{eq}}(T)\) is the equilibrium dimensionless polar order parameter and \(P_{\mathrm{scale}}\) converts it to \(\mu\mathrm{C\,cm^{-2}}\). The factor \(P_{\mathrm{scale}}\) does not enter the thermodynamic free energy, but is fitted together with the thermodynamic parameters using the polarization constraints described below. In the \(N_F\) regime, we use $P_s(172^\circ\mathrm C)\simeq 3.5~\mu\mathrm{C\,cm^{-2}}$, while in the ferroelectric smectic regime the polarization is constrained to remain close to $5~\mu\mathrm{C\,cm^{-2}}$ at \(165^\circ\mathrm C\), \(140^\circ\mathrm C\), and \(100^\circ\mathrm C\). These values are used as representative fitting anchors for the observed polarization magnitude and its approximately constant behavior in the ferroelectric smectic range, rather than as four independent high-precision experimental measurements.

\paragraph{Parameter constraints and fitting objective.}
The fitting is based on the reduced fixed-wave-number free energy \eqref{eq:reduced-energy}. To suppress unrealistically rapid low-temperature growth of the orientational and polar order parameters, we use
\begin{equation*}
    f_{\mathrm{fit}}=f+\frac{D_Q}{6}S^6+\frac{D_P}{8}p^8,\qquad D_Q>0,\quad D_P>0,\label{eq:calibration-energy}
\end{equation*}
where \(D_Q\) and \(D_P\) are higher-order stabilization coefficients for the orientational and polar order parameters, respectively. These terms suppress excessive low-temperature growth of the order parameters without introducing additional order parameters. All branch free energies used in the fitting are evaluated from \(f_{\mathrm{fit}}\).

We fix $C_Q=C_P=C_\psi=1$ as normalization conventions. Since the dimensionless smectic amplitude \(\rho\) has no independently fixed experimental scale, we additionally use $\rho(140^\circ\mathrm C)\simeq1$ to fix its numerical normalization. The remaining parameters are either fitted independently or derived from imposed transition conditions.

For the \(I\to N\) transition, we prescribe a small nematic jump \(S_{IN}=0.05\) at the experimental transition temperature \(T_{IN}=263^\circ\mathrm C\), so that the transition is represented as weakly first-order. The coexistence conditions
\[
    \partial_S f_{\mathrm{fit}}(S_{IN};T_{IN})=0, \qquad f_{\mathrm{fit}}(S_{IN};T_{IN})= f_{\mathrm{fit}}(0;T_{IN})
\]
give
\begin{equation}
    B_Q=3C_QS_{IN}+9D_QS_{IN}^3,\qquad T_Q^*=T_{IN}-\frac{C_QS_{IN}^2/3+(3/2)D_QS_{IN}^4}{a_Q}.\label{eq:derived-nematic-calibration}
\end{equation}
Hence \(B_Q\) and \(T_Q^*\) are derived from \(a_Q\) and \(D_Q\), rather than fitted independently.

For the continuous \(N\to N_F\) transition at \(T_{NF}=185^\circ\mathrm C\), the effective polar coefficient \(r_F\) defined in Eq.~\eqref{eq:near-N-polar-coefficients} satisfies \(r_F(T_{NF})=0\), with a positive effective quartic coefficient for the polar mode. The first condition determines
\begin{equation}
    T_P^*=T_{NF}-\frac{4\lambda}{3a_P}S_N^{\mathrm{fit}}(T_{NF}),\label{eq:derived-polar-calibration}
\end{equation}
where \(S_N^{\mathrm{fit}}(T_{NF})\) is the equilibrium nematic order parameter obtained by minimizing \(f_{\mathrm{fit}}(S,0,0,0;T_{NF})\) on the nematic branch. Thus \(T_P^*\) is also derived rather than fitted independently.

For each parameter set considered during the fitting, the nonzero order parameters of each branch retained in the experimental fitting are re-minimized at the temperatures entering the fit. The corresponding branch free energies are evaluated from \(f_{\rm fit}\) after full relaxation of the nonzero order parameters within each retained phase. Thus the fitting uses fully relaxed thermodynamic branches rather than prescribed empirical functions for \(S(T)\), \(p(T)\), \(\rho(T)\), or \(\theta(T)\). Within this restricted branch set, the equilibrium phase is identified as the branch with the lowest fully relaxed free energy.

The four experimental transitions enter the fitting in different ways. The \(I\to N\) and \(N\to N_F\) temperatures are imposed through \eqref{eq:derived-nematic-calibration} and \eqref{eq:derived-polar-calibration}, respectively. The \(N_F\to\SmA_F\) transition is fitted through the crossing of the fully relaxed \(N_F\) and \(\SmA_F\) branch free energies near \(167^\circ\mathrm C\). Equivalently, defining
\[
    \Delta F_{AF}(T)=F_{\SmAF}(T)-F_{N_F}(T),
\]
the fit requires \(\Delta F_{AF}\) to vanish near the experimental transition temperature, with the \(N_F\) branch stable above the transition and the \(\SmA_F\) branch stable below it. A small but finite discontinuity in the order parameters is also required to retain the weakly first-order character.

The \(\SmA_F\to\SmC_F\) transition is fitted as a continuous tilt instability of the \(\SmA_F\) branch. Denoting the relaxed quadratic and quartic tilt coefficients by \(r_C^{AF}\) and \(u_C^{AF}\), respectively, we require
\[
    r_C^{AF}(115^\circ\mathrm C)\simeq0,\qquad u_C^{AF}(115^\circ\mathrm C)>0,
\]
with \(r_C^{AF}>0\) above and \(r_C^{AF}<0\) below the transition. We also verify that no lower-energy finite-tilt state exists above the transition.

To combine these different fitting conditions in a single numerical objective, each condition is written as a dimensionless residual,
\[
    r_{b,j}=\frac{X_{b,j}^{\mathrm{mod}}-X_{b,j}^{\mathrm{tar}}}{s_{b,j}},
\]
where \(b\) labels a group of related fitting conditions, \(j\) labels an individual condition within that group, \(X_{b,j}^{\mathrm{mod}}\) is the model prediction, \(X_{b,j}^{\mathrm{tar}}\) is the corresponding experimental or imposed target, and \(s_{b,j}\) is the experimental uncertainty or a characteristic normalization scale. Inequality conditions, such as the required phase ordering across a transition, contribute to the residual only when they are violated.

For example, for a birefringence data point at temperature \(T_i\),
\[
    X_{Q,i}^{\mathrm{mod}}=n_{\mathrm{scale}}S_{\mathrm{eq}}(T_i),\qquad X_{Q,i}^{\mathrm{tar}}=\Delta n_i^{\mathrm{exp}},\qquad r_{Q,i}^{(\Delta n)}=\frac{n_{\mathrm{scale}}S_{\mathrm{eq}}(T_i)-\Delta n_i^{\mathrm{exp}}}{s_{\Delta n,i}},
\]
where \(s_{\Delta n,i}\) is the normalization scale for the \(i\)-th birefringence data point.

The residuals are grouped into five sectors: \(R_Q\) collects the orientational constraints, including the birefringence residuals and the imposed \(I\to N\) conditions; \(R_P\) collects the polar-ordering constraints for the \(N\to N_F\) transition; \(R_A\) collects the smectic-amplitude constraints for the \(N_F\to \SmA_F\) transition; \(R_{\rm pol}\) collects the spontaneous-polarization constraints; and \(R_C\) collects the tilt constraints for the \(\SmA_F\to \SmC_F\) transition. If \(N_b\) denotes the number of residuals in group \(b\), the final joint fit minimizes
\begin{equation}
    \mathcal J=\sum_b\frac{\omega_b^2}{N_b}\sum_{j=1}^{N_b} r_{b,j}^2,\qquad b\in\{Q,P,A,\mathrm{pol},C\},\label{eq:fitting-objective}
\end{equation}
with $(\omega_Q,\omega_P,\omega_A,\omega_{\mathrm{pol}},\omega_C)=(1,\;1,\;1.2,\;3.5,\;1.5)$. Different weights are assigned to the residual groups to balance their relative influence in the joint fitting.

\paragraph{Numerical fitting and final verification.}
Because the thermodynamic parameters are strongly coupled, the fit is initialized sequentially by adjusting the orientational \((a_Q,D_Q)\), polar \((a_P,B_P,\lambda,D_P)\), smectic \((a_\psi,T_\psi^*,B_\psi,g_1)\), and tilt \((g_1,g_P,g_2)\) sectors in turn. The derived quantities \(B_Q\), \(T_Q^*\), and \(T_P^*\) are updated throughout the procedure, and \(\chi_P\) is refined together with the polarization constraints.

This staged procedure is used only to generate a robust starting point in the coupled parameter space. In the final step, all independently fitted thermodynamic parameters, 
\[
    a_Q,\ D_Q,\ a_P,\ B_P,\ \lambda,\ D_P,\  a_\psi,\ T_\psi^*,\ B_\psi,\ \chi_P,\  g_1,\ g_P,\ g_2,
\]
together with \(P_{\mathrm{scale}}\), are released in a joint bounded nonlinear least-squares fit using the objective \eqref{eq:fitting-objective}. The derived quantities \(B_Q\), \(T_Q^*\), and \(T_P^*\) are recomputed at every evaluation and are never treated as independent fitting variables.

Since the free energy is invariant under \(p\mapsto-p\), only \(p\ge0\) is required. The equilibrium minimizations are performed over
\begin{equation*}
    0\le S\le1.25,\qquad 0\le p\le1.80,\qquad 0\le\rho\le2.80,\qquad 0\le\theta\le65^\circ.
\end{equation*}
These bounds define only the numerical search domain and are not used to define the phases. The fitted equilibrium solutions remain away from the upper bounds.

After the final joint fit, all branches retained in the experimental fitting are recomputed over the full temperature range and compared through their fully relaxed free energies. The \(I\to N\) and \(N\to N_F\) transition temperatures remain fixed by the imposed conditions above, while the \(N_F\to\SmA_F\) and \(\SmA_F\to\SmC_F\) transitions are located from the branch crossing \(F_{\SmAF}=F_{N_F}\) and the condition \(r_C^{AF}=0\), respectively. The resulting transition temperatures and transition types are reported in Table~\ref{tab:transition_character}.


The resulting thermodynamic parameters are listed in Table~\ref{tab:fit_parameters}.
\begin{table}[H]
\centering
\small
\renewcommand{\arraystretch}{1.22}
\setlength{\tabcolsep}{4pt}
\caption{Thermodynamic parameter set obtained from the experimental fitting.}
\label{tab:fit_parameters}

\begin{tabular*}{\textwidth}{@{\extracolsep{\fill}}cccccc@{}}
\toprule
Parameter & Value & Status & Parameter & Value & Status \\
\midrule
\(T_Q^*\)    & \(262.909\)  & Derived & \(T_P^*\)   & \(184.875\)  & Derived \\
\(a_Q\)      & \(0.009227\) & Fitted  & \(a_P\)     & \(0.010616\) & Fitted  \\
\(B_Q\)      & \(0.151049\) & Derived & \(B_P\)     & \(0.500071\) & Fitted  \\
\(D_Q\)      & \(0.932307\) & Fitted  & \(D_P\)     & \(1.726282\) & Fitted  \\
\midrule
\(T_\psi^*\) & \(143.238\)  & Fitted  & \(\lambda\) & \(0.001360\) & Fitted  \\
\(a_\psi\)   & \(0.016296\) & Fitted  & \(\chi_P\)  & \(0.124711\) & Fitted  \\
\(B_\psi\)   & \(-0.617828\)& Fitted  & \(g_1\)     & \(0.352151\) & Fitted  \\
\(g_P\)      & \(0.136866\) & Fitted  & \(g_2\)     & \(0.414002\) & Fitted  \\
\bottomrule
\end{tabular*}
\end{table}
The smectic layer spacing is reconstructed separately after the thermodynamic parameter set has been fixed. Introducing $c(T)=\sin\theta(T)$, we use the lowest-order even analytic representation
\begin{equation}
    q_{\mathrm{obs}}(T)=q_A+c_2c^2(T)+c_4c^4(T),\label{eq:q-observable-reconstruction}
\end{equation}
and define
\begin{equation}
    d(T)=\frac{2\pi}{q_{\mathrm{obs}}(T)}.\label{eq:d-observable-reconstruction}
\end{equation}
Here \(q_{\mathrm{obs}}(T)\) is an effective wave number used only to reconstruct the measured layer spacing and is distinct from the fixed thermodynamic reference wave number \(q_0\). The coefficients \(q_A\), \(c_2\), and \(c_4\) are determined by least-squares fitting to the temperature-resolved X-ray layer-spacing data after the thermodynamic parameters have been fixed. Consequently, \eqref{eq:q-observable-reconstruction}--\eqref{eq:d-observable-reconstruction} do not affect the branch free energies, equilibrium phase selection, or transition temperatures.

\section*{Acknowledgment}
L.Z. was supported by the National Natural Science Foundation of China (No. 12225102, T2321001, and 12288101) and the National Key Research and Development Program of China  2024YFA0919500. Y.H. was supported by the National Natural Science Foundation of China (Grant No. 1260011477), the Young Elite Scientists Sponsorship Program by the Beijing Association for Science and Technology, and the Young Faculty Development Fund of the School of Mathematics at Renmin University of China.
\bibliographystyle{ws-m3as}
\bibliography{references}
\end{document}